\pdfoutput=1
\documentclass[preprint]{elsarticle}
\usepackage[utf8]{inputenc}
\usepackage[T1]{fontenc}

\usepackage{amsthm, amsmath, amssymb, amsxtra}

\usepackage{hyperref}
\hypersetup{
	linktoc=all,
	colorlinks,
	citecolor=black,
	filecolor=black,
	linkcolor=black,
	urlcolor=black
}

\usepackage{graphicx}
\usepackage{color}
\usepackage{verbatim}
\usepackage{bbm}

\usepackage{dirtytalk}

\usepackage{mathtools}
\usepackage{xifthen}
\usepackage[toc,page]{appendix}

\include{HeadersAndCommands}

\journal{Theoretical Computer Science}

\usepackage{ifthen}

\begin{document}
\begin{frontmatter}

\title{General Cops and Robbers Games with randomness}
\author[uot]{Fr\'{e}d\'{e}ric Simard\corref{cor2}}
\ead{fsima063@uottawa.ca}

\author[ul]{Jos\'{e}e Desharnais\corref{cor1}}
\ead{josee.desharnais@ift.ulaval.ca}

\author[ul]{Fran\c{c}ois Laviolette\corref{cor2}}
\ead{francois.laviolette@ift.ulaval.ca}

%\cortext[cor1]{Principal corresponding author}
%\cortext[cor2]{Corresponding author}

\address[uot]{School of Electrical Engineering and Computer Science, University of Ottawa, Ottawa, ON, Canada}
\address[ul]{Department of Computer Science and Software Engineering, Universit\'e Laval, Qu\'ebec, QC, Canada}

\begin{abstract}

Cops and Robbers games have been studied for the last few decades in computer science and mathematics. As in general pursuit evasion games, pursuers (cops) seek to capture evaders (robbers); however, players move in turn and are constrained to move on a discrete structure, usually a graph, and know the exact location of their opponent. In 2017, Bonato and MacGillivray \cite{Bonatoa} presented a general characterization of Cops and Robbers games in order for them to be globally studied. However, their model doesn't cover cases where stochastic events  may occur, such as the robbers moving in a random fashion. In this paper  we present a novel model with stochastic elements that we call a Generalized Probabilistic Cops and Robbers game (GPCR). A typical such game is one where the robber  moves according to a probabilistic distribution, either because she is rather lost or drunk than evading, or because she is a robot. We present results to solve GPCR games, thus enabling one to study properties relating to the optimal strategies in large classes of Cops and Robbers games. Some classic Cops and Robbers games properties are also extended.
\end{abstract}

\end{frontmatter}
%\note{
%	Doit composer les highlights 
	%Highlights are a short collection of bullet points that convey the core findings of the article. Highlights are optional and should be submitted in a separate editable file in the online submission system. Please use 'Highlights' in the file name and include 3 to 5 bullet points (maximum 85 characters, including spaces, per bullet point). You can view example Highlights on our information site.
%}

\tableofcontents
\newcommand{\jd}[1]{\todo{jd: #1}}
%\newcommand{\rouge}[1]{\textcolor{red}{Arbitre 1: #1}}
%\begin{itemize}%\addtolength{\itemsep}{-4pt}
%% \item {If ..., then ...}
%\item The paper is fairly long; is there a way to shorten the discussion of the many examples of the paper? Perhaps describe one or two examples in depth and then just list the others.
%\item 
%It would be useful to understand the context, if any, of the game of Zombies and Survivors game within the papers’s framework. This is a one-player stochastic variant of Cops and Robbers, where the zombies (cops) always move directly toward the survivor (robber). It would be useful to give a (brief) discussion on this variant.
%% \item The general consensus is to capitalize “Cops and Robbers.”
%\item  I would suggest avoid using bullet items in formal writing and replace
%them by enumerated lists.
%% \item 
%%  It would be better to use $\subseteq$ throughout rather than $\subset$ .
%% \item  
%%  Avoid informal notation such as $\exists$ and $\Leftrightarrow$. I would also suggest omitting the less rigorous := in favor of =.
%\item 
% It would improve the presentation if the author’s refrain from beginning sentences with a symbol. See Definition 2.1 for this and later examples.
%\end{itemize}

%!TEX root = ms.tex
\definecolor{darkross}{rgb}{0.016,0.404,0.667}
\newcommand{\modifjo}[1]{\textcolor{darkross}{#1}}
\definecolor{darkred}{RGB}{198,51,51}
\newcommand{\modiffre}[1]{\textcolor{blue}{#1}}

\section{Introduction}
\newcommand{\red}[1]{\textcolor{red}{[#1]}}

Cops and Robbers games have been studied as examples of discrete-time pursuit games on graphs since the publication of Quilliot's doctoral thesis \cite{Quilliot1978f} in 1978 and, independently, Nowakowski and Winkler's article \cite{Nowakowski1983} in 1983. Both monographs describe a turn-based game in which a lone cop pursues a robber on the vertices of a graph. The game evolves in discrete time and with perfect information. The cop wins if he eventually shares the same vertex as the robber's, otherwise, if the play continues indefinitely, the latter wins. A given graph is \win{} if the cop  has a winning strategy: for any possible move the robber makes, the cop has an answer that leads him to eventually catch the robber (in finite time). As there is no tie, it is always true that  one player has a (deterministic) winning strategy.

Since the first exposition of the game of Cop and Robber, many variants have emerged. Aigner and Fromme \cite{Aigner1984a} notably presented in 1984 the \emph{cop number}: it is the minimal number of cops required on a graph to capture a robber. Since then, more alternatives have been described, each one modifying one game parameter or more such as the speed of the players, the radius of capture of the cops, etc. We refer to Bonato and Nowakowski's book \cite{Bonato2011f} for a comprehensive description of these different formulations. The survey on  guaranteed graph searching problems by Fomin and Thilikos \cite{Fomin2008} is also a great reference on the subject. In graph searching games, the objective is to capture a fugitive on a graph. The problems in which the object is always found are called guaranteed. 
%tiré de Fomin2008: " where the searchers have to guarantee the capture without any probabilistic assumption on the game behavior. " et plus loin " to rescue the explorer independent of his actions. In particular, the search strategy should be successful even if the explorer tries to avoid meeting with searchers."

In 2017 Bonato and MacGillivray \cite{Bonatoa} presented a first generalization of \coprob{} games that encompasses the majority of the variants described previously. Indeed, all two-player, turn-based, discrete-time, pursuit games of perfect information on graphs in which both players play optimally are contained in Bonato and MacGillivray's model. As such, this model encompasses all pursuit games deemed \emph{combinatorial} (we refer to Conway's book \emph{On Numbers and Games} \cite{Conway1976} for an introduction on the subject of combinatorial games). Those games include the set of turn-based, perfect information, games played on a discrete structure without any randomness. 
%{The classical example of combinatorial games is Chess. By opposition, Stratego \cite{Wikipedia2015}, although it is deterministic and discrete, is not considered combinatorial because it is played with imperfect information. Moreover, games of chance such as poker are not considered because of the stochasticity involved. Combinatorial games include no random events.}

Recently, some researchers such as Pra\l{}at and Kehagias \cite{Kehagias2012}, Komarov and Winkler \cite{Komarov2013b} and Simard et al. \cite{Simard2015} described a game, called the Cop and Drunk Robber game, in which the robber walks  in a random fashion: each of her movements is described by a uniform random walk on the vertices of the graph. In general, this strategy is suboptimal. Since this particular game cannot be described by Bonato and MacGillivray's model, it appears natural to seek to extend their framework to integrate games with random events.

There has also been a recent push towards more game theoretic approaches to modeling \coprob{} games, notably by Konstantinidis, Kehagias and others (see for example \cite{Konstantinidis2017,KEHAGIAS2013100,Kehagias2018,KEHAGIAS201725,KONSTANTINIDIS201648}). Our paper can be considered more in line with this way of treating \coprob{} games than more traditional approaches. 
%\modiffre{However, we did not make use of those previous research since they did not correspond to our goals. Either the authors modelled too specific games or used approaches (such as seeking Nash equilibria) that were too specific to their context. Notice that \cite{Kehagias2018} is also a generalization of Cops and Robbers games, however the extension proposed is in increasing the number of players instead of seeking more general descriptions of the rules of the game. }

This paper thus presents a model of \coprob{} games that is more general than that of Bonato and MacGillivray. The main objective of this model is to incorporate games such as the Cop and Drunk Robber game. The probabilistic nature of this game leads to define a framework different from the one of Bonato and MacGillivray.

%\note{Changer le titre de section}
%\subsection{Solving  \coprob{} games}

In \coprob{} games, one is generally interested in the question of \emph{solving} a game. This question is universal to game theory where one defines a \emph{solution concept} such as the \emph{Nash Equilibrium}. In \coprob{} games, often-times the cops' point of view is adopted and one seeks to determine whether it is feasible, and if so how, for them to capture the robbers. In stochastic \coprob{} games, one can generalize the question to a quantitative scale of success: what is the (best) probability  for the cops  to capture the robbers, and which strategy reflects it. One can also ask the dual question of what would be the minimal number of cops required in order to capture the robbers with some probability. In deterministic games, this graph parameter is known as the \emph{cop number}.

One can note that many solutions of \coprob{} games share the same structure, and this is reflected in the fact that they can be solved with a recursive expression. Indeed, Nowakowski and Winkler \cite{Nowakowski1983} in 1983 presented a preorder relation on vertices, writing $x\preceq_n y$ when  the cop has a winning strategy  in at most $n$ moves if positioned on vertex $y$, while the robber is on vertex $x$.  An important aspect of this relation $\preceq_n$ is that it can be computed recursively and thus leads to a polynomial time algorithm to compute its values, as well as the strategy of the cop. This relation was extended $20$ years later by Hahn and MacGillivray \cite{Hahn2006} 
in order to solve games of $k$ cops by letting players move on the graph's strong product.
 Clarke and MacGillivray \cite{Clarke2012}  have also defined a characterization of $k$-cop-win graphs through a dismantling strategy and studied the algorithmic complexity of the problem. 
For a fixed $k$ the problem can be resolved in polynomial time with degree $2k+2$. On a related note, Kinnersley \cite{Kinnersley2015} proved that it is \exptc{} to determine whether the cop-number of a graph $G$ is less than some integer $k$ when both $G$ and $k$ are part of the input.  This shows that Clarke and MacGillivray's result is somehow optimal.
%This suggests that solving cops and robbers games incurs high algorithmic cost and thus careful analysis of the complexity of our resolution method is presented in this paper. 

In games with stochastic components, such order relations can be generalized by considering  the probability of capture, as is done in a recent paper  about the \emph{Optimal Search Path} (OSP) problem \cite{Simard2015}. A recursion $w_n(x,y)$ is defined: it represents the probability that a cop standing on vertex $y$ captures the robber, positioned on vertex $x$, in at most $n$ steps. This relation, defined on the Cop and Drunk Robber game \cite{Kehagias2012,Komarov2013b,Simard2015}, is analogous to Nowakowski and Winkler's $x\preceq_n y$ and is slightly more general as it enables  to model the robber's random movement. 
 One can wonder up to what point  the relation $w_n$ can be extended while preserving its polynomial nature. Theorem~\ref{thm:copwinthm} and Proposition~\ref{prop:compabswn}  give an answer to this question.
 %\added[id=fs]{In practice, this relation will be defined on a possibly large structure, it will be deemed polynomial in regards to this object.}

This paper is divided as follows. Section \ref{sec:absmodel}  presents our model of  \coprob{} games, the $w_n$ recursion along with some complexity results, notably, on $w_n$. Stationarity results on $w_n$ are also included. Since most \coprob{} games are played on graphs, another formulation of our model is presented on such a structure in Section \ref{sec:concrgames}. We conclude in Section \ref{sec:conclusion}.

%!TEX root = ms.tex

\section{An abstract \coprob{} game}\label{sec:absmodel}
%attention il y a une note à ce sujet plus loin, où on parle de produit cartésien
% Subsection \ref{subsec:wneq} presents
%the $w_n$ recursion. Subsection \ref{subsec:compcomplex} presents 
%some complexity results, notably on the relation $w_n$. Stationarity results on $w_n$  follow in subsection \ref{subsec:station}. 
We now present a general model of \emph{Probabilistic} \coprob{} games; it is  played with perfect information, is turn-based starting with the cops,  and takes place on a discrete structure. 
%In the following definition, we give the ingredients of a specific game. This is sometimes called an arena~\cite{8,9}, a game being the conjunction of a specific graph/arena together with  rules for winning, definition of plays etc. However, since two  different games only differ by their arenas, we use the appellation \emph{game}, hereby  following the graph theoretic style. 
From each state/configuration of the game, after choosing their actions, the cops and robbers will jump to a state according to their transition matrices, denoted $\tr$ and $\tc$. These matrices may encode  probabilistic behaviours: $\tc(s,a,s')$\footnote{The notation $T(s,a,s')$ refers to a 
%https://www.cs.ubc.ca/~murphyk/Bayes/pomdp.html
\emph{transition matrix} view. In this way, it corresponds to annotating the edge $[s,s']$ of the transition system with an action $a$ and a positive value, the probability. In the Markov Decision Processes (MDP) community, it is also written $T_a(s,s')$, or  $\mathbb{P}(s'| s,a)$. } is interpreted as the probability that the cop,  starting in $s$ and playing action $a$, will arrive in $s'$.

\newcommand{\init}{i_0}
\begin{definition}\label{def:genabspurgame}
A \emph{Generalized Probabilistic \coprob{}} game (GPCR)  is played by two players, the \emph{cop} team and the \emph{robber} team. It  is given by the following tuple 

\begin{align}
\game 
&=
\left(
S, \init, F, A, \tc, \tr
\right),
\end{align}
satisfying
\begin{enumerate}

\item $S = \stc \times \str \times \sto$, the non-empty finite set of states representing the possible configurations of the game. The sets $\stc$ and $\str$ hold the possible cops and robbers positions while $\sto$ may contain other relevant information  (like whose turn it is).
\item $\init\in S$ is the  initial state.
\item $F\subseteq S$ is the set of final (winning) states for the cops.
\item $A = \ac\cup \ar$, with $\ac$ and $\ar$  the non-empty, finite sets of actions of the cops and robbers, respectively.
\item $\tc : S\times \ac \times S \rightarrow [0,1]$ is a \emph{transition function} for the cops, that is, 
$$\textstyle\sum_{s'\in S}\tc(s,a,s')\in \se{0,1} \mbox{ for all $s\in S$ and } a\in \ac.$$
When the sum is 1, we say that $a$ is playable in $s$, and we write $\ac(s)$  for the set of playable actions for the cops at state $s\in S$. 
Furthermore, $\tc$ also satisfies
\begin{itemize}%\addtolength{\itemsep}{-4pt}
\item for all $s\in S$, $\ac(s)\neq\emptyset$
\item if $s\in F$, then $\tc(s,a,s)=1$ for all action $a\in\ac$; hence $\tc(s,a,s')=0$ for all $s'\neq s$.%  
\end{itemize}

%Action $a$ can be played from state $s$ if and only if this value is $1$. 
\item $\tr$ is a transition function for  the robbers, similar to $\tc$. $\ar(s)$ is the set of playable actions by the robbers in state $s\in S$.
\end{enumerate}
A \emph{play} of $\game$ is an infinite sequence  $\init a_0s_1a_1s_2a_2\dots \in (S\ac S\ar)^\omega$ of states and playable actions of $\game$ that alternates the \emph{moves}\ of $\cops\!$ and $\robbers$.  It thus satisfies $\tc(s_j,a_j,s_{j+1})>0$ for $j = 0,2,4,\dots $ and $\tr(s_j,a_j,s_{j+1})>0$ for $j = 1,3,5,\dots $.  The cops win whenever a final state $s\in F$ is encountered, otherwise the robbers win. A turn
is a subsequence %$s_{2i}a_{2i}s_{2i+1}a_{2i+1}$ 
of two moves, starting from  $\cops$. 
We also consider finite plays and we write $\game_n$ for the game where plays are finite with $n$  (complete) turns.
\end{definition}
An equivalent formulation for $\tc$, and sometimes more handy, is to rather define $\tc(s,a)$ as a distribution on $S$,  for  an action $a$ playable in $s$. 
%That is, $\tc(s,a) :   \cl{P}(S)\rightarrow [0,1]$, 
The correspondance is $\tc(s,a)( X) =\sum_{s'\in X} \tc(s,a,s') $ for  $X\subseteq S$. For example, the second condition of the fifth item in the preceding definition could have been stated $\tc(s,a)=\dirac{s}$, where $\dirac{s}$ is the Dirac distribution on an element $s$, that is, $\dirac s$ has value 1 on $\{s\}$, and is 0 elsewhere.

A play progresses as follows: from a state $s$, the cops choose an action $\cac\in \acs{s}$, which results in a new state $s'$, randomly chosen according to distribution $\tc(s,\cac)$; then the robbers play an action $\rac\in \ars{s'}$, which results in  the next state $s''$, drawn with probability $\tr(s', \rac, s'')$. Once a final state is reached, the players are forced to stay in the same state. Notice that one could record whose turn it is in the third  component of the states: $\sto=\{\cops,\robbers\}$. However,  this doubles the state set and complexifies the definition of the transition function. In most games, it is more intuitive to define the rules for movement independently of when this transition will be taken, like in chess.

We  sometimes use the notation $s_{\x}$, for $\x\in \se{\cops, \robbers, \objects}$ to denote the projection of a state $s\in S$ on the set $S_{\x}$. The set $\sto$ is rarely used in the current section, but will be valuable further on, such as in Example~\ref{ex:dynamicgraphgame} on dynamic graphs whose structures vary with time. 

In what follows, we write $\dist{B}$ as the set of discrete distributions on a set $B$ and $\unif{B}\in \dist{B}$ for the discrete uniform distribution on the same set.

% In order to simplify some results, such as Theorem~\ref{thm:copwinthm}, we assume $I$ contains only one element. This assumption is made without much loss of generality, as will be observed in the coming examples.

Most of the example games we will describe will be between a single cop and a single robber, even if the definition specifies a cop team and a robber team.  The usual way of presenting the positions of the cop team  is with a single vertex in the strong product  of each member's possible territory. 
\begin{comment}
%\deleted{The same can be said of the robbers. From individual positions of each team members, one can  write, for example,  $\stc$ as a cartesian product of sets $\stc = \prod_{i=1}^k S_i^k$, where $S_i$ is the territory on which cop $i$ moves.} %Thus, we can always simulate $k$ cops playing on $\game$, that we consider a single team. 
%\red{Si on ne veut pas enlever, trouver une meilleure place/motivation pour en parler}
%\deleted{In the next definitions, we assume $k>0$ cops are playing on $\game$. }
\end{comment}

%\begin{definition}\label{def:abspurplay}
%Let $\game$ be a \coprob{} game. A play $\cl{H}$ of $\game$ is a possibly infinite sequence  $s_0a_0s_1a_1s_2a_2\dots \in (S\ac S\ar)^\omega$ of states and actions of $\game$ that alternates the moves of $\cops$ and $\robbers$. It thus satisfies $\tc(s_i,a_i,s_{i+1})>0$ for $i = 0,2,4,\dots $ and $\tr(s_i,a_i,s_{i+1})>0$ for $i = 1,3,5,\dots $.  It ends whenever a final state $s\in F$ is encountered,  in which case the cops have won and the robbers have lost. If $\cl{H}$ is infinite, then the robbers have won. \end{definition}

\subsection{Encoding of known games and processes, stochastic or not}
We now describe a few known games, following the structure of Definition~\ref{def:genabspurgame}. The first one is a typical, deterministic example of a \coprob{} game.
We say a game is \emph{deterministic} when both distributions defined by $\tc$ and $\tr$ are concentrated on a single point, in other words if $\tc(s,a)$ and $\tr(s,a)$ are Dirac for all $s\in S$ and $a\in A$. The  reader can safely skip this section.

\begin{example}[\bf Classic Cop and Robber game]\label{ex:classiccrgame}\rm
Let $G=(V,E)$ be a finite graph. In this game, both players play alone and walk on the vertices of the graph, successively choosing their next moves among their neighbourhoods. The final states are those in which both players share a vertex, in which case the cop wins. The tricky part for encoding this game is that in their first move, the cop and the robber can choose whatever vertices they want, so the rule of moving differs at the first move from the rest of the play. So we let $\ic, \ir\notin V$ be two elements that will serve as starting points for the cop and the robber. Because the first moves are  chosen in turn, the set of states $S$ below  must contain  states in $V\times \{\ir\}$, which can only be reached after the cop's first move, but before the robber's.  To simplify $S$, we include states that will not be reached, and this will be governed by the transition functions. The different sets are:
\begin{align*}
\init &= (\ic, \ir) \\
S &= (\se{\ic}\cup V)\times (\se{\ir}\cup V)\\
F &= \se{(x,x)\in V^2}\\
\ac &= V\\
\ar &= V.
\end{align*}
Let $(c,r)\in S$, $x\in V$, and actions $c'\in \ac$ and $r'\in \ar$. We define:
\begin{align*}
\tc((c,r), c', (x,r)) 
&=
\begin{cases}
1, &\mbox{ if } x=c' \mbox{ and } c = \ic \mbox{ or } c'\in N[c],\\
0, &\mbox{ otherwise.}
\end{cases}\\
\tr((c,r), r', (c,x))
&=
\begin{cases}
1, &\mbox{ if } x=r' \mbox{ and } r=\ir \mbox{ or } r'\in N[r],\\
0, &\mbox{ otherwise.}
\end{cases}
\end{align*}
Thus, for state $(c,r)\in S\setminus \{\init\}$, the playable action set is $\acs{c,r}=N[c]$. Similarly, for the robber we get $\ars{c,r} = N[r]$. Because a play  starts with the cop, it is  not required to specify the condition $c\neq \ic$ in  function $\tr$. Similarly,  is it not necessary to make a special  case of state $c=r$, since the play ends anyway. 
%The functions $\tc$ and $\tr$ both encode the sets of available actions for both players. From these functions, one can deduce the sets of actions from specific states. For example, 
\end{example}

The stochasticity of Definition~\ref{def:genabspurgame} is motivated by the following example, called the Cop and Drunk Robber game. It is rather similar to the one just presented except that the robber  moves randomly on the vertices of the graph.

\begin{example}[\bf Cop and Drunk Robber game]\label{ex:copdrunkrobber}\rm \label{ex:drunk}
From the preceding example, only the robber's transition function $\tr$ is modified, the rest stays the same.  Let $(c,r)\in S$ and $r'\in \ar$. The robber's transition function is then:
\begin{align*}
\tr((c,r), r')
&=
\begin{cases}
\dirac{(c,r')}, &\mbox{ if } r=\ir, \\
\unif{\{c\}\times N[r]}, &\mbox{ otherwise. }
\end{cases}
%then $\tr((c,r), r') : \se{c}\times V \rightarrow [0,1]$ is defined from \added{the distribution $D((c,r),r') :   \cl{P}(V)\rightarrow [0,1]$}:
%\begin{align*}
%D((c,r), r')
%&:=
%\begin{cases}
%\dirac{r'}, &\mbox{ if } r=\ir,\\
%\unif{N[r]}, &\mbox{ otherwise.}
%\end{cases}
\end{align*}
%This distribution  is on $\str$, but its extension to $S$ is straightforward by observing that the robber does not change the cop's position on the graph. 
The robber, after the first move, moves uniformly randomly on her  neighbourhood, which amounts to ignoring her action $r'\in \ar$.
One could also restrict her actions by $\ars{s} = \se{1}$
when $s\in S\setminus \{\init\}$. 
\end{example}

In the Cop and Drunk Robber game, the robber moves according to a uniform distribution on her neighbourhood. Varying her transition function could represent various scenarios. For example, the robber's probability of ending on a vertex $r'$ from vertex $r$ could  depend on the distance between $r$ and $r'$.

In addition to the Cop and Drunk Robber game itself, a recent paper by Simard et al.~\cite{Simard2015} presented a variant of this game in which the robber can evade capture. The main difference between these games is that the cop may not catch the robber even when standing on the same vertex. This game is presented in the next example.

\begin{example}[\bf Cop and Drunk Defending Robber]\label{ex:copdrunkdefrobber}\rm 
The game's main structure is again similar to that of Example~\ref{ex:classiccrgame}, but we need a jail to simulate the catch of the robber, $\jail\notin V$. The initial state is the same, and we have:
\begin{align*}
\init &= (\ic, \ir) \\
S &= (\se{\ic}\cup V)\times (\se{\ir}\cup V) \cup \se{(\jail, \jail)}\\
F &=\se{(\jail, \jail)}.
\end{align*}
When  players do not meet, they  move on $G$ as before. Yet, when the cop steps on the same vertex $v$ as the robber, there is a probability $p(v)$ the robber gets captured, where  $p : V\rightarrow [0,1]$. For $(c,r)\not\in F$, the robber's transition function is then:
\begin{align*}
\tr((c,r), r')
&=
\begin{cases}
\dirac{(c,r')}, &\mbox{ if } r=\ir, \\
\unif{\{c\}\times N[r]}, &\mbox{ if } c\neq r\mbox{ and } r\neq \ir,\\
D_{r}, &\mbox{ if } c=r\mbox{ and } r\neq \ir,
\end{cases}\\
\mbox{ where }
D_{r}(x)
&= 
\begin{cases}
\frac{1-p(r)}{|N[r]|}, &\mbox{ if } x\in \se{c}\times N[r]\mbox{ and } c = r, \\%}{x\in N[r]},\\
p(r), &\mbox{ if } x=(\jail, \jail).
\end{cases}
\end{align*}
When the cop steps on the robber's vertex ($c=r$), at the end of his turn, the next move for the robber follows the distribution $D_r$. The robber is caught by the cop with probability $p(r)$, bringing the play in a final state, otherwise she proceeds as expected: the target state is chosen uniformly randomly in the robber's neighbourhood. Variations of this game could be defined through different  distributions for $\tr((c,r), r')$ with $c\neq r$. Likewise, in $D_r$, the factor $\frac{1}{|N[r]|}$ could be replaced with any distribution on ${N[r]}$.
\end{example}

We now present the Cop and Fast Robber game with surveillance zone as first formulated in Marcoux \cite{Marcoux}. This example is reconsidered further on in Section \ref{sec:concrgames}. Chalopin et al.\  also studied a game of Cop and Fast Robber with the aim of characterizing graph classes~\cite{Chalopin2011}.

\begin{example}[\bf Cop and Fast Robber]\label{ex:copfastrobber}\rm
This game is similar to the classic one (Example~\ref{ex:classiccrgame}) except that the robber is not limited to a single transition. It has been studied by Fomin et al.~\cite{FominGKNS10}. We present a variation where the cop can capture the robber when she appears in his watch zone, even in the middle of a path movement. This watch zone can simulate the use of a weapon by the cop. The states will now contain, in addition to both players' positions, the set of vertices watched by the cop. We assume here that  the cop's watch zone is his neighbourhood, as in Marcoux \cite{Marcoux};  Fomin et al.'s version is retrieved with a watch zone consisting of a single vertex, the cop's position. In the initial state, the cop's watch zone is empty since the robber cannot be captured before her first step. We again use a jail state $\jail\notin V$. When both players find themselves there, the game ends and the robber has lost. Hence, we let:
\begin{align*}
\init &= (\ic, \emptyset, \ir) \mbox{ with } \ic, \ir \notin V,\\
F &= \se{(\jail, \emptyset, \jail)},\\
S &= 
\left(
\se{(\ic, \emptyset)} \cup \se{(c, N[c])\mid c\in V}
\right)
\times 
\left(
\se{\ir}\cup V
\right)
\cup F
%S &:= I\cup \se{(c, N[c], r) \mid c,r\in V} \cup F\mbox{\red{meme probleme que classique}}
\end{align*}
%The cop's transition function encodes his watch zone that is determined after his action. 
Let $(c,C,r)\in S$ be the current state and $c'\in N[c]$ an action of the cop. Here is the cop's transition function, for $(c,C,r)\not\in F$:
\begin{align*}
\tc((c,C,r), c')
&=
\begin{cases}
\dirac{(c', N[c'], r)}, &\mbox{ if } c=\ic\mbox{ and } c'\in V \mbox{ or }\\
			&\mbox{ if } c\in V\mbox{ and } c'\in N[c],\\
0, &\mbox{ otherwise.}
\end{cases}
\end{align*}
As in the classic game, the cop can jump to any vertex in his first move; after that he moves in the neighbourhood of his current position.  His watch zone then changes   to $N[c']$. We  use $C$ as watch zone in this definition to emphasize the fact that it does not influence the cop's next state. On her turn,  on vertex $r_1\in V$, the robber's action consists in choosing a path $\pi=(r_1, r_2, \dots, r_n)$ of finite length $n>0$, that is, $[r_i,r_{i+1}]$ is an edge in $E$ for each $i=1,2, \dots$. The robber's transition function is:
\begin{align*}
\tr((c,C, r_1), \pi)
&=
\begin{cases}
\dirac{(c, C, r_n)}, &\mbox{ if } r_1=\ir \mbox{ and } r_n \in V\setminus N[c], \mbox{ or}\\
		 &\mbox{ if } r_1\in V \mbox{ and } r_i\notin C\mbox{, for all } \;2\leq i\leq n,\\
\dirac{(\jail, \emptyset, \jail)}, &\mbox{ otherwise.}
\end{cases}
\end{align*}
The robber is thus ensured to reach her destination $r_n$ provided that she never crosses the cop's watch zone on her path $\pi$. If this happens, then the robber is taken to the jail state $(\jail, \emptyset, \jail)$. 

In Section \ref{sec:concrgames}, we present this  game again, but with the possibility for the robber to evade capture.
\end{example}

Hence, because of Definition~\ref{def:genabspurgame}'s rather general description, it is possible to encode a great variety of random events resulting from the cops' or the robbers' actions. In the following example, we encode a simple inhomogeneous Markov Chain by forgetting the notions of cop. This makes the example fairly degenerate but it also shows the generality of Definition~\ref{def:genabspurgame}.

 \begin{example}[\bf Finite Markov chain]\label{ex:finiteMarkovchain}\rm
 A Markov chain is a sequence of random variables $X_0,X_1,\dots$ on a space $E$, having the Markov property. So we can assume that the evolution is given by an initial distribution $q$ on $E$ and a family of matrices $M_0,M_1,\dots$, where $M_i(s,s')$ is the probability that $X_{i+1}=s'$ given that $X_i = s$.  We can encode it  as a GPCR game from Definition~\ref{def:genabspurgame}.  In previous examples, we have ignored the third component of states, $\sto$, but here we can ignore one of the players sets, like $\str$; equivalently, we can assume a single state for the robber and  no effect by $\tr$. We define
 \begin{align*}
  \init&\notin E\\
 S &= I \cup (E\times \mathbb N)\\
 F&=\emptyset\\
 A&=\se{1}\\
 \tc(i_0,1, (e,0)) &=q(e)\\
 \tc((e,j),1,(e',j+1))
 &=M_j(e,e').
 \end{align*}
 Since the action of the player has no influence on the progress of the game, it is natural to define $A$ as a singleton. Technically, a  play alternates between the moves of cops and robbers, so it is a sequence $i_0 1 (e_0,0) 1 (e_0,0) 1 (e_1,1) 1 (e_1,1)  \dots$; the repetitions reflect the fact that the robber has no effect. If we ignore the useless information of such a play, we obtain a sequence $i_0 e_0 e_1 e_2 \dots$,  which is just a walk in the Markov chain (and the robber wins). Another way to write down this model would have been to let the two players play similarly, with $\tr=\tc$, but the states would then have to be triplets, and the initial state would force a less simple encoding. 
 \end{example}

Similarly, we can encode a finite state Markov Decision Process (MDP) with reachability objectives~\cite{Puterman2014} with Definition~\ref{def:genabspurgame}. % Such an MDP can be naturally encoded by giving a reward of $1$ when the states to be reached are entered while the others give zero reward. If, in the  MDP, the set of states $F$ is reachable, then the optimal value is 1, otherwise it is 0. 
The encoding will satisfy that the optimal value of the MDP is 1 if the cops wins, otherwise it is 0, and the robber wins.

The probabilistic Zombies and Survivors game on graphs \cite{Bonato2016APV} can also be viewed as a GPCR game, one in which only the robbers play optimally. It models a situation in which a single robber (the survivor) tries to escape a set of cops (the zombies). However, the cops have to choose their initial vertices at random and, on each turn, choose randomly among the set of vertices that minimize the distance to the robber. 
%\begin{example}[\bf{Zombies and Survivors game}]\rm
%Let us reuse the same sets of Example \ref{ex:classiccrgame}, the Classic Cop and Robber game, where the cops' set of positions $V$ is replaced by $V^l$, for some $l>0$ the amount of cops chasing the robber. We simply modify the cops' transition function $\tc$. From the initial state $\ic$, $\tc((\ic,\ir), a) = \unif{V^l \times \se{\ir}}$ for any action $a\in V^l$. Then, from any other cops position $c\in V^l$ and action $a\in V^l$, $\tc((c,r),a) = \unif{N^* \times \se{r}}$, where $N^*\subseteq N_{G^l}(c)$ is the subset of neighbours of $c$ that minimize the distance to the robber. 
%\end{example}

\subsection{Strategies}
 A deterministic (or pure)  strategy is a function  that prescribes to a player which action to play on each possible game history. Some strategies are better than others;   we will be interested in the probability of winning for the cops, which will be attained by following a strategy. Ultimately, we are interested in memoryless strategies, that is, those that only depend on the present state, and not on the previous moves; nevertheless, we need to define more general strategies as well.
 \begin{definition}
Let $\game$ be a game. A  history on $\game$ is an initial fragment of a play on $\game$ ending in a state.  $H_\game$ is the set  of histories on $\game$.\vspace{-2mm}
\begin{itemize}\addtolength{\itemsep}{-6pt}
\item the set of \emph{general strategies} is  $\Omega^\mathrm{g} = \{\sigma: H_\game \rightarrow A\}$.
\item  the set of  \emph{memoryless strategies} is $\Omega=\{\sigma: S\rightarrow A\}$.
\item the set of \emph{finite horizon strategies} is $\Omega^{\mathrm {f}}=\{\sigma: (S\times {\mathbb N}) \rightarrow A\}$.
%, satisfying an time ellapsing property: $\forall s\in S, 
\end{itemize}
\end{definition}
 A finite horizon strategy  counts the number of turns remaining, and it is otherwise memoryless. A finite horizon strategy is conveniently defined on $\game$ but it is  actually played on $\game_n$, hence the following definition of how such a strategy is followed.   At turn 0 of $h$ (histories $\init$ and $\init a_0 s_1$), there are $n$ turns remaining, so $\sigma$ is evaluated with $n$ on the second coordinate of its argument; at turn 1 (histories $\init a_0 s_1 a_1 s_2$ and $\init a_0 s_1 a_1 s_2 a_2 s_3$),  there are $n-1$ turns remaining. 

\begin{definition}
Let $h=\init a_0 s_1 a_1 s_2 a_2 s_3\dots$  be a (finite or infinite) play of $\game$.
\begin{itemize}\addtolength{\itemsep}{-6pt}
\item $h$ follows a general strategy $\sigma\in\Omega^{\mathrm{g}}$ for the cops if  for all $j=0,2,4,\dots$ we have $a_j = \sigma(a_0 s_1 a_1 s_2 a_2 s_3\dots s_j)$. Similarly for the robbers.
\item $h$ follows a memoryless strategy $\sigma\in\strcset$ for the cops if  for all $j=0,2,4,\dots$ we have $a_j = \sigma(s_j)$. Similarly for the robbers. 
\item $h$ follows a finite horizon strategy  $\sigma\in\strcsetfini$ on $\game_n$ for the cops if  for $j=0,2,4,\dots, 2n$ we have $a_j = \sigma(s_j,n-\frac{j}{2}))$.  
\item $h$ follows a finite horizon strategy $\sigma\in\strrsetfini$ on $\game_n$ for the robbers if  $j=1,3,5,\dots, 2n+1$ we have $a_j = \sigma(s_j,n-\frac{j-1}{2})$.  
\end{itemize}
%We write $\strcset$ and $\strrset$ for the sets of deterministic and memoryless strategies of the cops and robbers, respectively.

\end{definition}

These strategies are all deterministic, or pure: a single action is chosen. Some papers consider \emph{mixed} or \emph{behavioral} strategies, where this choice is randomized. 
%\todo{les strategies mixed sont des distributions sur les strategies deterministes. les distributions sur les actions sont des behavioral strategies} 
% rep JD: j'ai ajusté pour englober les 2
This is unnecessary in our setting because, as is well known in perfect information games, among all optimal strategies, there is always a pure one. We will come back to this when we study optimal strategies later on.

We now present an example where the optimal strategy for the infinite game is memoryless (only depends on the states), but, for any finite horizon game $\game_n$, it is a finite horizon strategy.

\begin{example}
This example is in the spirit of the  Cop and Drunk Robber game, presented in Example~\ref{ex:drunk}.  As in this example, the cop moves on his neighbourhood  and so does the robber, who cannot choose her action, as before, but the difference with Example~\ref{ex:drunk} is that the robber's movement is not uniform. The graph is a cycle of length 5. The robber moves clockwise with probability 0.9, 
 and counterclockwise with probability 0.1. If the cop is at distance 1 of the robber at his turn, of course he wins in this turn. Otherwise, the cop is at distance 2, more specifically at \emph{clockwise distance}  2 or 3. Let us focus on states $s$ where this clockwise distance is 2 (from the cop to the robber). On the long term, the cop's best choice is to move counterclockwise. However, if  only one turn remains, the best move for the cop is the clockwise move because then with probability $0.1$, the robber will jump to his position, whereas the probability of winning is zero in the counterclockwise direction. So the best strategy $\sigma$ for $\game_n$ satisfies $\sigma(s,n)\neq \sigma(s,1)$ in such a state $s$, for $n>1$, hence it is not memoryless.  Indeed, for example, $\sigma(s,2)\neq \sigma(s,1)$ because the probability of catching the robbers by playing counterclockwise when 2 turns remain  is $0.9$, and it is $0.19$ by playing clockwise  ($0.1$ in one move of the robber plus $0.09$ in two moves).
 % This example can be generalized to cycles of length $2k+1$, where when there are $k$ turns left or less on a state from which the clockwise distance to the robber is $k$ or less, the best move is clockwise; otherwise it is counterclockwise. \modiffre{The counterclockwise distance is $n-2$ and in the best case both players reduce their distance by two on each turn. Thus, the cop needs at least $\lceil \frac{n-2}{2}\rceil$ turns to capture the robber moving counterclockwise and $\sigma(s,n) = \sigma(s,m)$ for $m\geq \lceil \frac{n-2}{2}\rceil$.}
\end{example}

%\begin{figure}
%\centering
%\input{ex_drunk_cycle}
%\caption{If the cop on vertex $C$ has one turn left to capture the robber on $R$, his best choice is to move clockwise.}
%\end{figure}

\subsection{Winning conditions in GPCR games}

In this section we  are interested in winning strategies for the cops, their probability of winning {in a given number $n$ of turns} {(that is, in $\game_n$) and their probability of winning without any limit on the number of turns (in $\game$).}

%on another hand, in the calculation of the number of turns that are necessary for them to win with a probability higher than some threshold.

%Assuming that both players play optimally, we define the probability  for the cop  to win in $n$ turns or less whatever the robber strategy:
%$$
% \pn := \p{\mbox{``capture in at most $n$ steps''}}
%$$
%Given a strategie $\strc$ %and $\strr$, for the cops and for the robbers, we  consider the probability that the robbers are captured in $n$ steps or less if the cops play optimally:
Given finite horizon strategies $\strc$ and $\strr$, for the cops and for the robbers, we  consider the probability that the robbers are captured in $n$ steps or less:
\begin{align*}
p_n(\strc,\strr) &:= 
\p{\mbox{\say{capture in at most $n$ steps}} \mid \strc,\strr}.\nonumber
%\pn(\strr) &= \max_{\strc\in \strcset} \p{\mbox{``capture in at most $n$ steps''} | \strc,\strr}\nonumber
\end{align*}
 Since the cops want to maximize this probability, and the robbers want to minimize it,  the probability  for the cops to win in $n$ turns or less (playing optimally), whatever the robbers strategy, is:
 \begin{align}
%\pn&:= \max_{\strc\in \strcset}\min_{\strr\in \strrset} \p{\ctime \leq n}.\\
\pn&:= 
\max_{\strc\in \strcsetfini}\min_{\strr\in \strrsetfini} p_n(\strc,\strr).
\label{eq:optgameval1}
\end{align}
This is in fact the \emph{value} of $\game_n$ in the sense of game theory. In game theory, the  value for $\game_n$ exists if
 \begin{align}
%\pn&:= \max_{\strc\in \strcset}\min_{\strr\in \strrset} \p{\ctime \leq n}.\\ 
\max_{\strc\in \strcset^\mathrm{g}}\min_{\strr\in \strrset^\mathrm{g}} p_n(\strc,\strr)
=
\min_{\strr\in \strrset^\mathrm{g}}\max_{\strc\in \strcset^\mathrm{g}} p_n(\strc,\strr)
.
\end{align}
% the maxmin in Equation~\ref{eq:optgameval1} is equal to the minmax.  
In our setting defining the  payoff function of a play   as 1 when the robbers are captured and 0 otherwise, we have, by Wal and Wessels~\cite{markovGames}, that  the game $\game_n$ has  value $\pn$.  That the restriction of $\pn$ to finite horizon strategies does achieves the value of $\game_n$ is given again by Wal and Wessels, who call such strategies Markov strategies. Finally, since $\game_n$ is finite and with perfect information, a standard game-theoretical argument \cite{osborne1994course} justifies that the optimal strategies are deterministic (or pure).

We say that the cops and the robbers play optimally in $\game_n$ if they each follow a strategy that yields probability $\pn$ for the cops to win. We will show later on, but it is also straightforward\footnote{This can be proven by induction, since for $n+1$ the cops can choose their optimal strategy for $n$ and simply do anything on the last turn.} from the definition, that $\pn$ is increasing in $n$; since it is moreover bounded by 1, the limit always exists and we will prove that is it equal to the value of $\game$.

Indeed, from a known result in Simple Stochastic Games (SSG), one can show that $\game$ has a value and that this value is achieved by a pair of optimal strategies that are deterministic (or pure) and memoryless. The argument is well known in the literature on SSGs, but requires a construction, so we leave it to Appendix \ref{sec:annex_ssg}. Thus,
let us write the value of game $\game$ as $\pg$, that is,
\begin{align}
\pg &=
\max_{\strc\in \Omega_c}
\min_{\strr\in \Omega_r}
\p{\mbox{\say{capture in a play}} \mid \strc, \strr},
\end{align}
and the equality still holds when the $\min$ and $\max$ operators are switched. This value is guaranteed by Theorem \ref{thm:ssgvalexist} \cite{condon1992complexity,shapley1953stochastic}.
In Proposition \ref{prop:epsoptimal}, we will show that the difference in the cop using a finite-horizon strategy in $\game_n$ and a memoryless one in $\game$ is negligible for a sufficiently large integer $n$.

Equation \eqref{eq:optgameval1} returns either $0$ or $1$ in deterministic games such as the Classic Cop and Robber game. We seek here to study games that can be stochastic, where $\pn$ can take any value in $[0,1]$. Thus, we adapt the usual definition of \emph{\win{}} to our broader model.

\begin{definition}\label{def:winningconditions}
Let $\game$ be a GPCR game. We say $\game$ is \vspace{-2,5mm}
\begin{itemize}\addtolength{\itemsep}{-6pt}
\item \emph{\cpwin{n}} if the cops can ensure a win with probability at least $p$ in at most $n$ turns, that is $\pn \geq p$;
\item 
 \emph{$p$-copwin} if it is \cpwin{n} for some $n\in\mathbb{N}$;
%asymptotically almost surely winning (a.a.s. winning), or \emph{almost surely \win{}}, 
\item \emph{almost surely copwin} if the cops can win {when they are allowed to play infinitely}, that is $\pg  %= \lim_{n\to\infty}\ptime{n}&
= 1$;
 \item 
  \emph{\win{}} if it is \cwin{n} for some $n\in\mathbb{N}$.  
\end{itemize}
\end{definition}
%\emph{\cpwin{n}} if the cops can ensure a win with probability at least $p$ in at most $n$ turns, that is:
%\begin{align*}
%\pn &\geq p.
%\end{align*}
%We say that $\game$ is \emph{$p$-copwin} if it is \cpwin{n} for some $n\in\mathbb{N}$, and that it is \emph{\win{}} if it is \cwin{n} for some $n\in\mathbb{N}$. Finally, we say a game is 
%%asymptotically almost surely winning (a.a.s. winning), or \emph{almost surely \win{}}, 
%almost surely copwin if the cops can win \added{in an unbounded number of turns}, that is:
%\begin{align*}
%\pg  %= \lim_{n\to\infty}\ptime{n}&
%= 1.
%\end{align*}
It is easy to see that when $\game$ corresponds to the Classic Cop and Robber game, as defined in Example~\ref{ex:classiccrgame}, this definition of copwin coincides with the classical one. In that sense, it can be considered as a generalization of the classical one, because in any copwin finite graph, the cop wins in at most $n=\abs{V(G)}^2$ turns.

%
%We observe that under certain conditions, the cops are ensured that their probability of winning is $1$. For example, if $\game$ is the Cop and Drunk Robber game (Example ~\ref{ex:copdrunkrobber}) played on a cycle, then the robber has a positive probability of choosing a suboptimal move on each vertex and hence the cop can expect to eventually capture her.

\begin{remark}
We will see in Proposition \ref{prop:epsoptimal} that $\lim_{n\to\infty}\pn = \pg$. Thus, if there exists $n$ such that $\pn >0$ and if all states reachable within {a finite number of moves} of the cop's optimal strategy are in the same strongly connected component,  then  $\pg = 1$. {Indeed, after $n$ turns, if the play is not over, the cops can go back to the configuration where $\pn>0$: the initial position that is proposed by the cops' strategy. In that state, the probability that the robbers have not been caught is at most $1-\pn$;  the probability that the robbers are not caught after $m$ repetition of this cycle is  at most $(1-\pn)^m$. It is thus zero at the limit.} This happens, for example, if $\pn >0$ and $\game$ is a strongly connected graph. However, we cannot, in general, claim that if $\pn>0$ after $n>\abs{S}$ turns have been played, then $\pg = 1$.
%If the first state $s_0$ selected by the cops from $\ic$ leads to $\pn >0$ for some $n>|S|$ and if $s_0$ is selected by the optimal strategy after $\ic$ has been left, then $\ptime{\infty} = 1$.
\end{remark}

We define a probabilistic analog to the cop number, $c(G)$, which is the minimal number of cops required on a graph $G$ in order for the cops to capture the robbers.  It is an important subject of research  in Classic \coprob{} games~\cite{Bonato2011f}, in particular relating to Meyniel's conjecture that  $c(G) \in \bigO{\sqrt{\abs{V(G)}}}$. Furthermore, one of the main areas of research on cops and robbers games that involve random events is the expected capture time of the robbers \cite{Komarov2013b,Komarov,Kehagias2012}. Thus, we further generalize the expected capture time of the robbers for any game $\game$. 

{Adding cops in a game  $\game$  is done in the natural way. The set of cops states $\stc$ is the cartesian product of the sets of single cop positions, and the transition function is updated so as to let all cops move in one step.}
\begin{definition}
The \pcopntext{n} $\pcopnsym{n}$ of a game $\game$ is the minimal number of cops required for the capture of the robbers in at most $n$ turns with probability at least $p$. In other words, $\pcopnsym{n}$ is the minimal number of cops required for a game $\game$ to be $\cpwin{n}$. The $p$-cop number, $\pcopsym = \pcopnsym{\infty}$,  is the minimal number of cops necessary for having $\pg\geq p$.

Let $T_\game^p$ be the random variable giving the number of turns required for the robbers to be captured with probability at least $p$ in $\game$ under optimal strategies. Then, the $p$-expected capture time of the robbers is $\e{T_\game^p}$. The \emph{expected capture time of the robbers} is $\e{T_\game^1}$.
\end{definition}

Since some of the optimal strategies of $\game$ are memoryless, we can turn the question of computing $\e{T_{\game}^p}$ into a question of computing an expected hitting time in a Markov chain. Let us write $\sigma_{\cops}^*$ ($\sigma_{\robbers}^*$) for the optimal strategy of the cops (robbers) in $\game$ and let $\cl{M}$ be the Markov chain such that for any state $s\in S$, it has two states $(s,\sigma_{\cops}^*(s))$ and $(s,\sigma_{\robbers}^*(s))$. Furthermore, let $M$ be its transition matrix, that is governed by the distributions $\tc(s,\sigma_{\cops}^*(s))$ and $\tr(s,\sigma_{\robbers}^*(s))$. Suppose $(X_n)_{n\geq 0}$ describes the stochastic process on $\cl{M}$ beginning at the initial state $\init$, then $T:= \frac 1 2 \min_{n\geq 0}(X_n\in F)$ is the hitting time of $F$ from $\init$. The expectation of $T$ is $\e{T_\game^1}$.
\subsection{Solving GPCR games}\label{subsec:wneq}

Similarly as with Bonato and MacGillivray's model, we define a method for solving GPCR games, that is, for computing the probability for the cops to capture the robbers in an optimal play, and the strategy to follow. This method takes the form of a recursion, defining the probability $w_n(s)$ that state $s$  leads to a final state in at most $n$ steps ($w$ is for \emph{winning} in the following theorem). This recursion gives a  strategy for the cops.

%Theorem~\ref{thm:copwinthm}, the main theorem of this section, justifies  the structure of Equation \eqref{eq:absgamewn}.% as it justifies this equation is indeed the robbers' probability of capture.
\begin{theorem}\label{thm:copwinthm}
Let $\game$ be a GPCR game, and let:
\renewcommand{\cac}{a}
\renewcommand{\rac}{a'}
\begin{align}\label{eq:absgamewn}
w_0&(s) :=
\begin{cases}
1, &\mbox{ if } s\in F,\\
0, &\mbox{ otherwise.}
\end{cases}\nonumber\\
w_n&(s) :=\nonumber\\&
\begin{cases}
\mbox{~}~1, &\mbox{ if } s\in F,\\
\displaystyle
\max_{\!\!\cac\in \scalebox{0.8}{$\acs{\!s'\!}\!\!$}} \sum_{s'\in S} \!\tc(s,\cac, s')\displaystyle
\min_{\!\!\!\!\rac\in \scalebox{0.8}{$\ars{\!s'\!}\!\!$}} \sum_{\scalebox{0.7}{$s''$}\in S} \hspace{-0.8mm}\tr(s', \rac, s''\hspace{-0.2mm}) w_{n-1}(s''\hspace{-0.1mm} ), \hspace{-3.3mm}\mbox{}
&\mbox{ otherwise.}
\end{cases}
\end{align}
 Then $w_n(s)$ gives the probability for the robbers to be captured in $n$ turns or less, given that both players play optimally, starting in state $s$.  Thus,
$$
w_n(\init) = \pn.
$$
This also says that  $\game$ is \cpwin{n} if and only if $w_n(\init) \geq p$.
For $(s,k)\in S\times\mathbb{N}$, let %$\sigma^*_\cops: S\times \mathbb{N}\to A $  as: 
$\sigma^*_\cops(s,k)$ be the $\argmax$ in place of $\max$ in Equation~\eqref{eq:absgamewn}.
%This is a finite horizon strategy and it is optimal in $\game_n$.
This defines finite horizon strategies that are optimal in $\game_n$\footnote{The argmax is not necessary unique.}.

\end{theorem}
The recursive part of $w_n$'s definition is as follows: to win, the cops must take the best action $a$; this leads them to state $s'$ with probability $\tc(s,a,s')$; from this state, the robbers must choose the  action $a'$ that will  give them  the smallest probability of being caught. Action $a'$ leads the robbers to state $s''$ with probability $\tr(s',a',s'')$ and then we multiply by the probability that the cops catch the robbers from this state, $w_{n-1}(s'')$. Since the cops want a high probability, a maximum is taken; it is the converse for the robbers.  The full equation  gives the expected probability of capture of the robbers by the cops when both players move optimally.

\begin{proof}
The proof is by induction on $n$. We prove that   $w_n(s)$ gives the probability for the robbers to be captured in $n$ turns or less, given that both players play optimally, starting in state $s$.
%for any state $s$ considered as initial state for the computation of $\pn$, we have $w_n(s) = \pn$.
Let $s$ be any state.  

If $n=0$, then the cops win if and only if $s\in F$, in which case,  by definition  we do have $w_0(s)=1$. Otherwise the robbers win and $w_0(s)=0$, as wanted.
 %both players stand still. If the robbers have ended their turn in a final state, i.e. if $s\in F$, then with probability $1$ the cops have won. Otherwise, the cops have lost.

If $n>0$, suppose the result holds for $n-1\leq k$ and let $s$ be the current state. If this state is final, then the robbers are caught in $n$ turns or less with probability $1$ and $w_n(s) =1$ as desired. Otherwise, let the cops, playing first, choose an action $\cac\in \acs{s}$, after what the next state $s'$ is drawn according to $\tc(s, \cac, s')$. Then, the robbers can choose an action $\rac \in \ars{s'}$, in which case the next state $s''$ will be drawn with probability $\tr(s', \rac, s'')$. By the induction hypothesis, we know a final state will be encountered in $n-1$ turns or less with probability $w_{n-1}(s'')$ starting from state $s''$. Thus, the probability the robbers are caught in $n$ turns or less by playing action $\rac$ after the cops have reached state $s'$ is given by:
\[
\sum_{s''\in S} \tr(s', \rac, s'')w_{n-1}(s'').
\]
Note that if $s'\in F$, this value is exactly $w_{n-1}(s')$, since by definition, we must have  $\tr(s', \rac, s'') = 1$ if $s''=s'$ and 0 otherwise.
The robbers wish to minimize this value among their set of available actions, which is possible since both sets $S$ and $\ar$ are finite. Hence, supposing  action $\cac\in\acs{s}$  has been chosen by the cops, the game stochastically transits to some other state $s'\in S$ with probability $\tc(s,\cac, s')$. Thus,  with probability 
\[
\sum_{s'\in S}\tc(s,\cac, s')
\min_{\rac\in \ars{s'}}
\sum_{s''\in S}
\tr(s', \rac, s'')
w_{n-1}(s''),
\]
the robbers are caught in at most $n$ turns from state  $s$, when the cops play action $\cac$. The cops want to maximize this value and, as for the robbers, this is possible because the considered sets are finite. Thus, the cops must play the action 
\[
\argmax_{\cac \in \acs{s}}
\sum_{s'\in S}\tc(s,\cac, s')
\min_{\rac\in \ars{s'}}
\sum_{s''\in S}
\tr(s', \rac, s'')
w_{n-1}(s'').
\]
The claim about $\sigma^*_\cops$ is straightforward from this result. The choices of actions at the initial state thus give the probability $w_n(\init)$. Because $\pn$ is, by definition, the probability of capture of the robbers in $n$ turns or  less when both players play optimally, we conclude that $w_n(\init) = \pn$.
\end{proof}

This result implies that the  $w_n$'s are probabilities that increase with $n$. In other words, we have the following corollary. 

\begin{corollary}\label{cor:mono}
For any $n\in\mathbb{N}, s\in S$ we have $0\leq w_n(s)\leq w_{n+1}(s)\leq 1$.
\end{corollary}

Note that there are many optimal strategies for the cops in $\game$, that is, stategies that have value $\pn$, but they are not all as efficient. Consider a game $\game_{n}$ where the robbers can be caught in $k<n$ turns with probability 1, and let $\sigma_k$ be an optimal strategy for $\game_k$. Then the strategy that stays idle for $n-k$ turns and then behave as prescribed by $\sigma_k$ is optimal, but not efficient, and it respects the argmax of Equation~\eqref{eq:absgamewn}. The next proposition shows how to define an efficient one.
 
\begin{proposition}\label{pr:sigma*N}
For each $N\in\N$, there exists an optimal strategy $\sigma_N^*$ with horizon $N$ that satisfies:
for all $s\in S$, if $w_{N_1}(s)=w_{N_2}(s)$ for $ N_1\leq N_2\leq N$, then $\sigma_{N}^*(s, N_1)=\sigma_{N}^*(s,N_2)$. {Similarly for the robbers.}
 \end{proposition}

\begin{proof}
%First assume that all strategy defined by Equation~\eqref{eq:absgamewn} chosen at horizon 
For any $(s,m)\in S\times \N$, we denote by $\mbox{\rm ACT}(s,m)$  the set of actions that achieve the maximum in Equation~\eqref{eq:absgamewn} for $w_m(s)$. We have proved in Theorem~\ref{thm:copwinthm} that any strategy satisfying $\sigma(s,m)\in\mbox{\rm ACT}(s,m)$ for all $(s,m)\in S\times \N$ is optimal.
Let us prove that if $w_{N_1}(s)=w_{N_1+1}(s)$ for $ N_1\in\N$, then $\mbox{\rm ACT}(s,N_1)\subseteq\mbox{\rm ACT}(s,N_1+1)$.
By contradiction let  $k$ be the smallest integer such that there is a state $s$ and an action $a\in \mbox{\rm ACT}(s, k)\setminus\mbox{\rm ACT}(s,k+1)$. By induction, the cops play action $a$ at time $k+1$, and then, with horizon $k$,  they  choose an optimal action in $\sigma_{k-1}^*$ that is also in $\sigma_{k}^*$ (possible by minimality of $k$) and so on until the last turn where they stay in place or whatever possible. This gives us a value of at least  $w_{k}(s)$ and, by definition, at most  $w_{k+1}(s)$. Since  $w_{k}(s)= w_{k+1}(s)$, the finite horizon strategy defined above is optimal. This is a contradiction since $a$ should be in $\mbox{\rm ACT}(s,k+1)$. Thus we obtain that if  $w_{N_1}(s)=w_{N_2}(s)$ for $ N_1\leq N_2\leq N$, then $\mbox{\rm ACT}(s,N_1)\subseteq\mbox{\rm ACT}(s,N_2)$, for all  $ N_1\leq N_2\leq N$. Hence the wanted strategy exists. {The argument is similar for the robbers.}
\end{proof}

Although $w_n(i_0)$ only gives the value of the game $\game_n$ with finite-horizon strategies, we can show that this relation, as a function of $n$, converges to the value of $\game$.

\begin{proposition}\label{prop:epsoptimal}
The value of $\game$ is $\lim_{n\to\infty}\pn $. Furthermore, the optimal strategies of $\game_n$ are $\epsilon$-optimal strategies of $\game$ for any $\epsilon>0$ and sufficiently large integer $n$.
\end{proposition}
\begin{proof}
From a previous argument, we know that some pair $(s_c,s_r)$ of optimal memoryless strategies for the cops and the robbers  yields a probability $\pg$ of winning for the cops. 
It holds that $\pn \leq \pg$, for any integer $n$, since the value of $\game_n$ can only be at most the value of $\game$. Recall that since $\pn$ is non-decreasing in $n$ and bounded above by $p_\game^*$, we have that $\lim_{n\to\infty}\pn \leq \pg$.

Now, let us play strategies $(s_c,s_r)$, chosen above, in the game $\game_n$ for any integer $n$. Consider the probability that the cops win in $\game_n$ when both players follow those strategies. These probabilities, for each $n$, form a sequence $(v_n)_{n\in \N}:=(v_1,v_2, \dots)$. This sequence is non-decreasing and bounded above by $p_\game^*$. 

Let $A_n$ be the event \say{there is a capture in at most $n$ turns under strategies $s_c$ and $s_r$}. Observe that $A_0\subseteq A_1 \subseteq \dots$ is a non-decreasing sequence. Thus, by the Monotone Convergence Theorem:
\begin{align*}
p_\game^* 
&= \p{\{h\mid \mbox{$h$ is a play following $s_c, s_r$ where cops win} \}}\\
&= \p{\cup_{i=0}^\infty A_i}\\
&= \lim_{n\to\infty}\p{A_n}\\
&= \lim_{n\to\infty}v_n.
\end{align*}
Thus, for any $\epsilon>0$ there exists an integer $N$ such that for all $n\geq N$, $p_\game^*-v_n < \epsilon$. But, we also have that $v_n\leq \pn$ for any integer $n$,  since $w_n(i_0)$ is the value of $\game_n$. Hence, it follows that $0< p_\game^* - \pn\leq p_\game^* - v_n = \abs{p_\game^*-v_n} <\epsilon $. This completes the proof.
\end{proof}

It is interesting to note that this theorem only applies if there are best strategies for the cops and robbers. In particular, it is not true if $\game$ is played on the {infinite graph of the following example.}
\begin{example}
Consider an \emph{infinite star graph} with a central vertex, from which paths of lengths $n$ are deployed, for every integer $n$, and consider the Classic \coprob{} game $\game$ on this graph with one cop and one robber. The best move for the cop is to start on the (infinitely branching) central vertex. Then whatever state the robber chooses, the cop will catch her in a finite number of turn, so this graph is copwin in the sense of Definition~\ref{def:winningconditions}. However this number of turn is unbounded, so when playing in $\game_n$, the robber can simply choose a vertex at distance greater than $n$; so the value of $\game_n$ is 0 for all $n$. The proof of the theorem fails in that case because,  the graph being infinite, there is no optimal strategy for the robber in $\game$. Whatever state the robber chooses, there is always a further state that would allow her to be captured in more turns, that is, there is always a better strategy.
\end{example}

Under certain conditions that will be further studied in Subsection \ref{subsec:station}, the $(w_n)_{n\in\mathbb{N}}$ sequence becomes constant. 

\begin{definition}
We say that $(w_n)_{n\in\mathbb{N}}$ is \emph{stationary} if there exists an integer $N\in \N$ such that $w_{n}(s) = w_{n+1}(s)$, for all $n>N$,  $s\in S$.
We   write $\sw$ for the stationary part of $(w_n)_{n\in\mathbb{N}}$.
\end{definition}

\begin{remark}\label{r:grandN}
It follows from the definition of $w_n$ that, if  for some $N$, $w_N(s)=w_{N+1}(s)$ for all $s\in S$, then $(w_n)_{n\in\mathbb{N}}$ is {stationary}  and $\sw$  starts at $n=N$ or less. 
\end{remark}

From Theorem \ref{thm:copwinthm}, we deduce  Theorem \ref{th:fixpointcor} that is more in line with traditional game theoretical arguments and show that in addition to the equality $\lim_{n\to\infty} w_n (\init) = \pg$ we can compute explicitly the optimal strategy of the cops in $\game$, from the limit of the $w_n$'s.
%First, given a non empty set of final states $F$, let us define $w_\infty$ the relation such that for any state $s\in S$: 
\begin{theorem}\label{th:fixpointcor}
\renewcommand{\cac}{a}
\renewcommand{\rac}{a'}
The (point-wise) limit  $w_\infty := \lim_{n\to\infty} w_n$ exists and it satisfies 
\begin{align}\label{eq:w_infini}
w_\infty&(s) =\nonumber\\&
\begin{cases}
\mbox{~}~1, &\mbox{ if } s\in F,\\
\displaystyle
\max_{\!\!\cac\in \scalebox{0.8}{$\acs{\!s'\!}\!\!$}} 
\sum_{s'\in S} \!\tc(s,\cac, s')\displaystyle
\min_{\!\!\!\!\rac\in \scalebox{0.8}{$\ars{\!s'\!}\!\!$}} 
\sum_{\scalebox{0.7}{$s''$}\in S} 
\hspace{-0.8mm}
\tr(s', \rac, s''\hspace{-0.2mm}) 
w_{\infty}(s''\hspace{-0.1mm} ), \hspace{-3.3mm}\mbox{}
&\mbox{ otherwise.}
\end{cases}
\end{align}
Moreover, the optimal (memoryless) strategy for the cops in $\game$, from any state $s$, can be retrieved by a cops' action for which the maximum of Equation~\eqref{eq:w_infini} is achieved.
\end{theorem}
\begin{proof}
\renewcommand{\cac}{a}
\renewcommand{\rac}{a'}
Let $L$ be the lattice of functions $S \rightarrow [0,1]$, ordered point-wise, with the null function as bottom element $\bot$. Equation~\eqref{eq:absgamewn} determines the following function ${\cal F} : L\rightarrow L$. For $f: S\to [0,1]$ and $s\in S$,
\begin{align}
{\cal F} (f) &(s) :=\nonumber
\\&
\begin{cases}
\mbox{~}~1, &\mbox{ if } s\in F,\\
\displaystyle
\max_{\!\cac\in \scalebox{0.8}{$\acs{s'\!}\!$}} \sum_{s'\in S} \!\tc(s,\cac, s')\displaystyle
\min_{\!\!\!\!\rac\in \scalebox{0.8}{$\ars{s'\!}\!$}} \sum_{\scalebox{0.7}{$s''$}\in S} \hspace{-0.8mm}\tr(s', \rac, s''\hspace{-0.2mm}) f(s''\hspace{-0.1mm} ), \hspace{-3.3mm}\mbox{}
&\mbox{ otherwise.}
\end{cases}\nonumber
\end{align}
From previous remarks, ${\cal F}$ is monotone increasing. Thus, we deduce from the Knaster-Tarski fixed point theorem \cite{granas2003}  that ${\cal F}$ has a least fixed point given by  $w_\infty :=\lim_{n\to \infty} {\cal F}^n(\bot)$. 
%Observe that ${\cal F}(w_\infty) = w_\infty$ so $w_\infty$ is a fixed point of ${\cal F}$. 
Furthermore, we have
${\cal F}(\bot) = w_0$ and  ${\cal F}(w_{n-1}) = w_n$, so ${\cal F}^{n+1}(\bot) = w_n$,  for all integer $n$, and thus $w_\infty = \lim_{n\to \infty} w_n$ and satisfies Equation~\eqref{eq:w_infini}. 

We showed in Theorem~\ref{thm:copwinthm} that $w_n(\init) = \pn$, and in Proposition \ref{prop:epsoptimal} that $\lim_{n\to\infty} \pn= \pg$. 
Consequently, $w_\infty(\init) = \pg$. Hence, $w_\infty(\init)$ is the probability that the cops capture the robbers when both teams play optimally. Similarly, one can show that $w_\infty(s)$ is the probability that, starting at $s$, the cops capture the robbers when both team play optimally.
%Thus, since ${\cal F}^n(\bot) (\init)\leq w_\infty(\init) \leq \pg$, it follows that $w_\infty(\init) = \pg$.
This, together with the fact that $w_\infty$ satisfies Equation~\eqref{eq:w_infini}, imply that the optimal strategy for the cop is coherent with an action  achieving the $\argmax$ operator in  place of the $\max$ operator in Equation~\eqref{eq:w_infini}. One cannot choose any such action because, for example, a temporary bad action, like staying idle, can give the same probability of winning than another action, but you can only choose it a finite number of times, which is incompatible with a memoryless strategy.
\end{proof}

\begin{remark}
Recall that we have $w_n(i_0) = \pn$ and that, by definition, it holds that $\pn = \min_{\strr\in \strrset^\mathrm{g}}\max_{\strc\in \strcset^\mathrm{g}} p_n(\strc,\strr)$. Thus, we could have defined $w_n(i_0)$ with switched operators $\min$ and $\max$. Then, we can deduce the optimal robbers strategies by flipping those operators and replacing the $\min$ operator by an $\argmin$ operator. This also holds in $w_\infty$.
\end{remark}

Now, with the help of Equation \eqref{eq:absgamewn} we can {generalize} the classic theorem of 
\coprob{} games. This is done in the next corollary. 
\begin{corollary}
Let $\game$ be a GPCR game. Then, $\game$ is \emph{\win{}} if and only if the sequence $(w_n)_{n\in\mathbb{N}}$ is stationary and
\[
\sw(\init) = 1.
\]
Moreover, the game is  \pcopwin{} if and only if the sequence is stationary and 
\[
\sw(\init) \geq p.
\]
If $\game$ is not $\pcopwin{}$ for any $p$, then the game is almost surely \win{} if and only if the sequence is not stationary and
\[
w_\infty(\init) =1.
\]
\end{corollary}

\begin{remark}\label{r:determstation}
If the GPCR game  $\game$ is deterministic, then $w_n(s)$ is $0$ or $1$ for any $n\in\mathbb{N}$ and $s\in S$. It therefore follows from monotonicity of $(w_n)_{n\in\mathbb{N}}$ (see Corollary~\ref{cor:mono}) and from Remark \ref{r:grandN} that the stationary part starts at some $N\leq \abs{S}$. Indeed, if $w_n\neq w_{n+1}$ there is at least one $s$ such that $w_n(s)=0$ and $w_{n+1}(s)=1$. This difference can be observed at most $\abs{S}$ times.
\end{remark}

The conditions under which  $(w_n)_{n\in\mathbb{N}}$ is stationary are presented in Proposition \ref{prop:wnstation}.

\subsection{The computational complexity of the \texorpdfstring{$w_n$}{wn} recursion}\label{sec:compwn}

We show a result on the algorithmic complexity of computing function $w_n$ (Equation \eqref{eq:absgamewn}). This function is computable with dynamic programming, yet it may require a high number of operations, especially as its complexity is function of the size of the state space. Recall that Equation \eqref{eq:absgamewn} was devised to be as general and efficient as possible. However, given the context of Definition~\ref{def:genabspurgame}, the best one can hope for its polynomial complexity in the size of the state and action spaces.
\begin{proposition}\label{prop:compabswn}
In the worst case and under a dynamic programming approach, computing $w_n$ requires $\bigO{n\abs{S}^3\max |\ac|\max |\ar|}$ operations, where $\max |\ac|$ is $\max_{s\in S}\abs{\ac(s)}$, similarly for $\max |\ar|$. The spatial complexity is $\bigO{n\abs{S}}$.
\end{proposition}
\begin{proof}
Let $a_n$ be the number of operations required for computing the recursion of $w_n$. Assume that computing probabilities $\tc$ and $\tr$ require unit cost. Clearly, $a_0=1$. In the worst case, when $n>0$, all elements of the sets $\ac$ and $\ar$ must be considered in order to ensure optimality of the actions chosen and thus $\textstyle\max |\ar|\max |\ac|$ operations are required. We always have that $\abs{S}>\abs{\se{s'\in S \mid \tc(s,\cac,s')>0}}$ and similarly for $\tr(s,\rac,s')$. Then, in the worst case, 
\begin{align*}
a_n &\leq \abs{S}^3\textstyle\max |\ar|\max |\ac| + a_{n-1}\\
&\leq n\abs{S}^3\textstyle\max |\ac|\max |\ar|+ 1,
\end{align*}
where we assumed that all values of $w_{n-1}$ were saved in memory for all $n-1$. Memorizing those values requires a spatial complexity of $\bigO{n\abs{S}}$ at most. The final complexity is thus $\bigO{n\abs{S}^3\max |\ac|\max |\ar|}$.
\end{proof}

Consequently, both spatial and temporal algorithmic complexities depend on the three  sets  $S$, $\ac$ and $\ar$. This suggests that these complexities may be high if the number of available actions is. One could imagine a game in which actions are paths,  resulting in exponential complexity in $\abs{S}$. Still, whenever $\ac\in \bigO{p(\abs{S})}$ and $\ar\in \bigO{q(\abs{S})}$ for some polynomials $p$ and $q$, then Equation \eqref{eq:absgamewn} is clearly computable in polynomial time in the size of $S$. 
Moreover, as we will see in Corollary~\ref{cor:wncomp}, $w_n$ does not have to be computed for all $n$ in order to determine if  the cops have a winning strategy or not, essentially, $n=\abs{S}$ suffices. 
In many studied cases, $|S|$ is itself polynomial in the size of the structure on which the game is played, leading each time to polynomial time algorithms for solving the game. 

\subsection{A stationarity result}\label{subsec:station}

In traditional games of \coprob{} where a relation $\preceq_n$ is defined (such as the classic game \cite{Nowakowski1983} and the game with $k$ cops \cite{Clarke2012}), it is useful to prove results on the convergence of the recursion $\preceq_n$. One demonstrates the relation becomes \emph{stationary}, that is, there exists a number $N\in \N$ such that for all integers $n>N$ and all pairs of vertices $(u,v)\in V^2$, if $u\preceq_n v$, then $u\preceq_{n+1}v$. One then writes $\preceq$ for the stationary part of the sequence, i.e. $\preceq = \preceq_N$. This result is vital for solving \coprob{} games as it ensures the relation $\preceq$ can be computed in finite time.

% \red{[cette phrase est sans lien avec l'explic qui suit...?]The question of the stationarity of relation $w_n$ is not trivial. } 
Contrary to the relation $\preceq_n$ found in deterministic Cops and Robbers games (such as the classic game in Example~\ref{ex:classiccrgame}), the relation $w_n$ does not always become stationary. For example, on the triangle $K_3$, with one cop and one robber, although it is \win{} in the classical sense, whenever one adds a probability of capture on the vertices, say $1/M$ for $M>0$, then after $n$ turns the cop will have captured the robber with probability only $1 - (1-\frac 1 M)^n$. Thus, after $n$ turns, the cop can only ensure a probability of capture strictly less than $1$, although he can clearly win with probability $p$ for any $p\in [0,1]$. In other words, a game may be almost surely copwin, but not \cwin{n} for any integer $n$. In the following proposition, we formulate and prove an upper bound on the minimal number of steps $n$ required to determine $\pg$, the probability of capture in an infinite game.

Recall that it does not hold in general that in a copwin graph (in the classical sense of one cop against one robber) every optimal strategy of the cop prevents him from visiting any vertex more than once~\cite{boyer2013cops}. Were this to be true, we could easily upper bound the capture time of the robber. However, we show in Lemma~\ref{lem:uniquestatesplay} that a milder version of this result holds for states, instead of simple cop position.

To have an intuition of why the following lemma is true, it is important to note that the condition of stationarity is a very strong one. The contraposition of the lemma may be more informative: the only way for $w_n$ to become stationary is that there is no loop possible in any play  following the optimal strategies of the players. An example of a graph where it  does not happen is a cycle of length 3, where the robbers have equal probability in both directions in every state. There are plays where the robbers are caught after an arbitrary large number of turns. On the other hand, an acyclic graph does induce stationarity for $w_n$. 
\begin{lemma}\label{lem:uniquestatesplay}
Suppose $(w_n(s))_{n\in\N}$ is stationary at  $N>0$ in a game $\game$, for a state $s$ and that the cops and robbers follow their optimal strategy, from Proposition~\ref{pr:sigma*N}.
Then, every winning play  (for the cops) from $s$  brings the cops in any given state at most once (at the end of a turn).
\end{lemma}
\begin{proof}
We prove the result for both the cops and robbers, that is, in a winning play where they follow their optimal strategy, none of them visit the same state twice at their turn.
Because of stationarity  and Proposition~\ref{prop:epsoptimal} the optimal strategy for the cops  in $\game$  is also optimal in $\game_{n}$, $n\geq N$. Suppose the lemma is false. Then there is a winning play $\pi$ (i.e., reaching $F$ in $N$ turns or less) from state $s$ containing a loop through a state $s_k$ that is thus reached twice by the same player in the play, the second time being at $s_l$, $k<l$ (with $k$ and $l$ having the same parity).  This play follows the optimal strategies of the players. None of the states of the loop are in $F$ by definition of a play. Consider the set $\Pi$ of plays $\pi_i$ that  start as $\pi$ until the first occurence of $s_l$,  follow the fragment  $s_l a_l s_{l+1}\ldots a_{k-1} s_k$ from $s_l$ for  $i$ times, and then continue as the fragment of $\pi$ after it exits $s_l=s_k$  the last time. These are plays (in particular, they are alternating between the players). All these plays are winning (one of them may be $\pi$), but infinitely many of them reach $F$ at a turn greater than $N$.   If we prove that these plays follow the  optimal strategies, this contradicts stationarity as the robbers are caught in more than $N$ turns in an infinite number of them, which implies that the value of $\game_{N+k}$ is strictly greater than the value of $\game_N$  for infinitely many $k$.

%robber has advantage in staying the longest possible in the loop (even forever), the exit of the loop is not her choice. Similarly, because the strategy is optimal, entering the loop is not a choice of the cop, as cutting up the loop would be more efficient. 
We do have that any play of $\Pi$ follows the optimal strategies. Indeed, since $\pi$ follows the optimal memoryless strategyies, everytime the play reaches state $s_k = s_l$, the same action is chosen for the player. In the first occurence, it leads to enter the loop, in the last one it leads to leave it. This happens when the action leads to a stochastic next state. 
\\
\end{proof}

Note that the lemma is not true if the robbers do not play well. Indeed consider the very simple deterministic game played on a cycle of length greater than 3; a robber can avoid capture indefinitiely by traveling  away from the cop. Then $(w_n(s)_{n\in \mathbb{N}}$ is stationnary  for every state $s$. Consider a play where the robber decides to stop after having traveled 8 times around the cycle. The play is winning for the cop, but even if the cop follows the optimal strategy, the same state is encountered 8 times.   

\begin{proposition}\label{prop:wnstation}
Let $\game$ be a  GPCR game and $s\in S$. Then,  the recursion $w_n$ defined by Equation \eqref{eq:absgamewn} is such that:
\begin{enumerate}
\item if  $w_{\abs{S}}(s) = 0$, then for every $k>0$, $w_{\abs{S}+k}(s) =0$;
%\item if for all state $s\in S$, $w_{\abs{S}+1}(s) = w_{\abs{S}}(s)\in (0,1]$, then for every $k>0$, $w_{\abs{S}+k}=w_{\abs{S}}$;
% \item if $w_{\abs{S}}\neq w_{\abs{S}+1}$, that is, there exists a state $s\in S$ such that $w_{\abs{S}+1}(s)>w_{\abs{S}}(s)$, then $(w_n)_{n\in\mathbb{N}}$ is not stationary, or equivalently, $w_n(s)< w_\infty(s)$ for all $n\in \mathbb{N}$, $n\geq \abs{S}$.
\item  if  $w_{\abs{S}+1}(s)>w_{\abs{S}}(s)$, then $(w_n(s))_{n\in\mathbb{N}}$ is not stationary.
\end{enumerate}

\end{proposition}
\begin{proof}
For the first claim, assume that $w_{\abs{S}+k}(s) > 0$. Then there is a  path  from state $s$ to a final state in $F$ that follows $\sigma^*_{\abs{S}+k}$ (and that has positive probability).  If this path is longer than $\abs{S}$ then it contains a repetition of at least one state $s'$, at turns, say, $m_1$, and $m_2$.
Consider the finite horizon strategy  that follows  $\sigma^*_{\abs{S}+k}$ for the first $m_1$ turns, and then follows   $\sigma^*_{\abs{S}+k -m_2}$, which is the strategy followed by $\sigma^*_{\abs{S}+k}$ when $s'$ of $\pi$ was encountered for the second time originally in $\pi$.
So removing from $\pi$ the subpath between $m_1$ and $m_2$, we obtain a shorter path that has positive value and that follows this strategy. 
%This strategy is general, but we know that there is a finite horizon strategy that achieves its value. 
%\modifjo{As in the previous proof, by Wal and Wessels~\cite{markovGames}, there is a finite horizon strategy that achieves the same value.} 
By continuing this procedure, we obtain a path of length $|S|$ or less and Claim 1 is proved. 

%For the second claim, assume $(w_n)_{n\in\N}$ is stationary from $N\in\N$. Then $w_N(s)= w_{N+k}(s)=w_\infty(s)$ for all $s\in S$ and $k\in\N$. 
% We first prove that any play that follows  strategy $\sigma_N^*$ (as defined in Proposition~\ref{pr:sigma*N}) cannot go through the same state twice with different choices of actions from that state.  Assume a play following  $\sigma_N^*$ meets state $s'$ twice, at turns   $m_1$ and $m_2$, $m_1< m_2< N,$ and that  
% $\sigma_N^*(s',N-m_1)\neq\sigma_N^*(s',N-m_2)$. 
%Then $w_{N-m_1}(s')>w_{N-m_2}(s')$ by Proposition~\ref{pr:sigma*N} and by monotonicity.  We follow the following (general) strategy on $s$ in game $\game_{N+(m_1-m_2)}$.  For the first $N$ turns, we follows  $\sigma_N^*$ as if the horizon was $N$ instead of $N+(m_1-m_2)$. We do that, 
%except on $(s',N-m_2)$ where we play action $\sigma_N^*(s',N-m_1)$, and then we follow  $\sigma_{N-m_1}^*$. For plays that that do not contain $(s', N-m_2)$, there are $m_1-m_2$ turns left, where we play anything. Let $p$ be the probability that a path go through $(s',N-m_2)$ in $\game_{N+(m_1-m_2)}$ when following $\sigma_N^*$ as describe above. By construction, $p>0$, and we therefore have
% \begin{align*}
%w_N(s) 
%&= w_{N+m_2}(s) \\
%&\geq (1\!-\!p) w_N(s) + p (w_{N\!-\!m_1}(s'))\\
%&=   (1\!-\!p) w_N(s) + p  (w_{N\!-\!m_2}(s')) +  p(w_{N\!-\!m_1}(s')\!-\!w_{N\!-\!m_2}(s'))\\
%&= w_N(s) +  p(w_{N\!-\!m_1}(s')\!-\!w_{N\!-\!m_2}(s'))\\
%&> w_N(s)
%\end{align*}
%a contradiction.  

%because  then there exists a finite horizon strategy having value at least equal to this general one.
From Lemma \ref{lem:uniquestatesplay}, if $(w_n(s))_{n\in\N}$ is stationary from $N$, there is no (positive, or winning) plays where the same state is encountered twice  in the $N$ first turns of $\game_{N}$ following $\sigma_{N}^*$. Now, suppose $N\geq |S|$. Thus, there is no repetition of states, which implies that for all $s\in S$, all paths that contribute to the value $w_N(s)$  are of length at most $|S|$, and the result follows.
\end{proof}

It is interesting to note the contrapositive of the second item in Proposition~\ref{prop:wnstation}, that if $(w_n(s))_{n\in \N}$ is stationary for some state $s$, then $w_{\abs{S}}(s) = w_{\abs{S}+1}(s)$. In other words,  the stationary part starts at most at turn $\abs{S}$. This result is \emph{by state}, so other states may not be stationnary.  Note however that we cannot deduce stationnarity from observing $w_{\abs{S}}(s) = w_{\abs{S}+1}(s)$ because the sequence may stay stable for a few turns and then be updated with a positive value.
Anyway, we can complete the algorithmic complexity presented in Proposition \ref{prop:compabswn}.
\begin{corollary}\label{cor:wncomp}
In the worst case, under a dynamic programming approach at most $\bigO{\abs{S}^4\max |\ac|\max |\ar|}$ operations are sufficient in order to determine whether $w_n$ is null, stationary equal to a number $p\in (0,1]$ or infinitely increasing.
\end{corollary}
\begin{proof}
The result follows from Proposition \ref{prop:wnstation} and \ref{prop:compabswn} by substituting $n$ for $\abs{S}$. For stationnarity, for example,  if $(w_n)_{n\in \N}$ is stationnary, then $(w_n(s))_{n\in \N}$ is stationnary for all $s\in S$, so we can conclude that $(w_n)_{n\in \N}$ is stationnary at $n=\abs{S}$.
\end{proof}

\subsection{Bonato and MacGillivray's generalized Cops and Robbers game}
This subsection is dedicated to our comparison with Bonato and MacGilli\-vray's generalized \coprob{} game \cite{Bonatoa}, which is another attempt at studying \coprob{} games in general forms. For the sake of self-containment, their model is transcribed here. This model is completely deterministic and thus is included as a special case of Definition \ref{def:genabspurgame}. 

Bonato and MacGillivray's game is presented in the following definition.
\begin{definition}[Bonato and MacGillivray's game]\label{def:bonatomacggame}
A discrete time process $\game$ is a generalized \coprob{} game if it satisfies the following rules :
\begin{enumerate}
\item Two players, \emph{pursuer} and \emph{evader} compete against each other.
\item There is perfect information.
\item There is a set $\cl{P}_P$ of admissible positions for the pursuer and a set $\cl{P}_E$ for the evader. The set of admissible positions of the game is the subset $\cl{P}\subseteq \cl{P}_P\times \cl{P}_E$ of positions that can be reached according to the rules of the game. The set of game states is the subset $\cl{S}\subseteq \cl{P}\times \se{P, E}$ such that $((p_P, q_E), X)\in \cl{S}$ if, when $X$ is the player next to play, the position $(p_P, q_E)$ can be reached by following the rules of the game.
\item For each game state and each player, there exists a non-empty set of allowed movements. Each movement leaves the other player's position unchanged. We write $\cl{A}_P(p_P, q_E)$ the set of allowed movements for the pursuer when the game state is $((p_P, q_E), P)$ and $\cl{A}_E(p_P, q_E)$ for the set of movements allowed to the evader when the game state is $((p_P, q_E), E)$.
\item The rules of the game specify how the game begins. Thus, there exists a set $\cl{I}\subseteq \cl{P}_P\times \cl{P}_E$ of admissible starting positions. We define $\cl{I}_P = \se{p_P : \exists\; q_e\in \cl{P}_E, (p_P, q_E)\in \cl{I}}$ and, for $p_P\in \cl{P}_P$, we define the set $\cl{I}_E(p_P) = \se{q_E\in \cl{P}_E : (p_P, q_E)\in \cl{I}}$. The game $\game$ starts with the pursuer choosing a starting position $p_P\in \cl{I}_P$ and then the evader choosing a starting position $q_E\in \cl{I}_E(p_P)$.
\item After both players have chosen their initial positions, the game unfolds alternatively with the pursuer moving first. Each player, on his turn, must choose an admissible action given the current state.
\item The rules of the game specify when the pursuer has captured the evader. In other words, there is a subset $\cl{F}$ of final positions. The pursuer wins $\game$ if, at any moment, the current position belongs to $\cl{F}$. The evader wins if his position never belongs to $\cl{F}$.
\end{enumerate}
Only \coprob{} games in which the set $\cl{P}$ is finite are considered. Games considered are played on a finite sequence of turns indexed by natural integers including $0$.
\end{definition}

We also present how the same authors defined an extension of the relation $\preceq_n$ of Nowakowski and Winkler \cite{Nowakowski1983} in order to solve the set of games characterized by their model.
\begin{definition}[Bonato and MacGillivray's $\preceq_n$]\label{def:bonatopreceq}
Let $\game$ be a \coprob{} game given by Definition \ref{def:bonatomacggame}. We let :
\begin{enumerate}
\item $q_E\preceq_0 p_P$ if and only if $(p_P, q_E)\in \cl{F}$.
\item Suppose that $\preceq_0, \preceq_1, \dots, \preceq_{i-1}$ have all been defined for some $i\geq 1$. Define $q_E\preceq_i p_P$ if $(p_P, q_E)\in \cl{F}$ or if $((p_P, q_E), E)\in \cl{S}$ and for all $x_E\in \cl{A}_E(p_P, q_E)$ either $(p_P, x_E)\in \cl{F}$ or there exists some $w_P\in \cl{A}_P(p_P, x_E)$ such that $x_E\preceq_j w_P$ for some $j<i$.
\end{enumerate}
By definition, $\preceq_i$ contains $\preceq_{i-1}$ for all $i\geq 1$. Since $\cl{P}_E\times \cl{P}_P$ are finite, there exists some $t$ such that $\preceq_t=\preceq_k$ for all $k\geq t$. We define $\preceq=\preceq_t$.
\end{definition}

Bonato and MacGillivray then use the relation defined in Definition~\ref{def:bonatopreceq} to show a necessary and sufficient condition for the existence of a winning strategy for the pursuer that is greatly similar to corresponding theorem of Nowakowski and Winkler \cite{Nowakowski1983}. 
%This theorem is presented here for the sake of completeness, its demonstration is however omitted as it can be recovered in their paper \cite{Bonatoa}.
\begin{theorem}[The \emph{\win} theorem of Bonato and MacGillivray]
The pursuer has a winning strategy in a game of \coprob{} characterized by Definition \ref{def:bonatomacggame} if and only if there exists some $p_P\in \cl{I}_P$ such that for all $q_E\in \cl{I}_E(p_P)$, either $(p_P, q_E)\in \cl{F}$ or there exists $w_P\in \cl{A}_P(p_P, q_E)$ such that $q_E\preceq w_P$.
\end{theorem}
It should be clear at this point that both Definitions \ref{def:genabspurgame} and \ref{def:bonatomacggame} describe alternative pursuit games of perfect information that unfold on discrete structures. Although the notation is different in both cases, it should also be clear that Bonato and MacGillivray's model is embedded in ours. The only difference between our formalism has to do with the initial states. Indeed, we only allow one initial state, $\init$, which is not the case in Definition \ref{def:bonatomacggame}. This does not cause any problem as it suffices to play one more turn in our model, or even to modify the first reachable states. In order to simplify what follows, we assume the set of initial states in both models are equivalent. We conclude that Equation \eqref{eq:absgamewn} should encode the relation $\preceq_n$ of Definition \ref{def:bonatopreceq}. Indeed, other than its deterministic character, the relation $\preceq_n$ is greatly similar to our recursion. Both relations are binary and recursive. Both share the same structure: a single case when $n=0$ in which both players may not make another move; a second case when $n>0$, but the current state is final; finally, a last case, again when $n>0$, when both players must choose an action that is optimal in the subsequent turns. Thus, we formally show how those two equations are related in the coming lines.

We first note that Equation \eqref{eq:absgamewn} can be simplified when following Bonato and MacGillvray's model. Since the component $s_{\objects}$ is not used in what follows, we simply write $(c,r)\in S$. Since the game is deterministic, we let players choose their next position directly. The recursion $w_n$ is thus given by :
\begin{align}\label{eq:detergenabswn}
w_0(c,r) &=1 \iff (c,r)\in F;\nonumber\\
w_n(c,r) &=
\begin{cases}
1, &\mbox{ if } (c,r)\in F;\\
\displaystyle
\max_{
c'\in \acs{c,r}}
\min_{
r'\in \ars{c',r}}w_{n-1}(c', r'), &\mbox{ otherwise.}
\end{cases}
\end{align}

The following theorem thus makes the connection between the two formalisms that are our model and that of Bonato and MacGillivray. In order to clarify the exposition, the relation $\preceq_n$ is written in our model, that of Definition \ref{def:genabspurgame}. Given the preceding remarks, this should incur no loss of generality. 

\begin{theorem}
Let the relation $\preceq_n$ be given by Definition \ref{def:bonatopreceq} and $w_n$ the recursion given by Equation \eqref{eq:absgamewn}. Assume $\game$ is a GPCR game given by Definition \ref{def:genabspurgame}, but following the specifications of Definition \ref{def:bonatomacggame}. Then, we have:
\begin{align}\label{eq:wncorpreceq}
w_n(c,r) =1 \iff \exists \; \cac \in \acs{c,r} : r\preceq_n c'.
\end{align}
\end{theorem}
\begin{proof}
First, observe that relation $\preceq_n$ compares the positions of the pursuer and the evader. These positions are encoded in the game states $S$ of our model. Moreover, the set of actions $\cl{A}$ defined in model \ref{def:bonatomacggame} are in fact restrictions of the set of actions from model \ref{def:genabspurgame}. Indeed, actions in $\cl{A}$ directly correspond to game positions, whereas we enable, in definition \ref{def:genabspurgame}, the action sets to be disjoint from the set of states. It is thus possible to define a game $\game$ that respects the hypotheses of Definition \ref{def:bonatomacggame} and where Expression \eqref{eq:wncorpreceq} is well-defined. A subtle difference between both formalisms has to do with the turn counters: in relation $w_n$ the cops are next to play while the robbers are to make their move in relation $\preceq_n$. This does not change the fact that cops play first in both games. Now, we prove the result by induction, similarly as in the proof of Proposition~\ref{prop:classwneq}.
\\\emph{Base case: $n=0$.} $w_0(c,r) = 1$ if and only if $(c,r)\in F$ and $(c,r)\in F$ if and only if $r\preceq_0 c$.
\\\emph{Induction step.} Assume the result holds for $n\leq k$ and let's show it for $n=k+1$. It holds that $w_{k+1}(c,r)=1$ if and only if $(c,r)\in F$, in which case $r\preceq_{k+1}c$ by definition, or there exists an action $c' \in \acs{c,r}$ for the cops such that no matter the response $r'\in \ars{c', r}$ of the robbers, we have $w_k(c', r') = 1$. By the induction hypothesis, we have $w_k(c', r')=1$ if and only if there exists an action $c'' \in \acs{c', r'}$ such that $r'\preceq_k c''$. Thus, if the cops play action $c'$, they position themselves on a state in which $r\preceq_{k+1} c'$. Conversely, assume there exists an action $c'\in \acs{c,r}$ such that $r\preceq_{k+1}c'$. Then, by definition, for all response $r'\in \ars{c', r}$ of the robbers there exists an action $c''\in \acs{c', r'}$ of the cops such that $r'\preceq_k c''$. In this case, by the induction hypothesis, we have $w_k(c', r')=1$. The cops play action $c'\in \acs{c,r}$, in which case we have $w_{k+1}(c,r)=1$.
\end{proof}

%!TEX root = ms.tex

\section{A concrete model of GPCR games}\label{sec:concrgames}

In this section we present a more concrete model of GPCR games that is closer to the usual definitions in the literature. Thus, we specify that  the game is played on a graph, without pointing out its particular shape. The actions of the players will correspond to paths as in the game of Cop and Fast Robber \cite{Marcoux}. The game presented in Definition~\ref{def:genabspurgame} is abstract because its sets do not depend on any precise structure and so neither does the algorithmic complexity of computing Equation \eqref{eq:absgamewn}. The point of reformulating Definition~\ref{def:genabspurgame} is to refine some results and formulate them in terms of the graph's structure.

\subsection{Definition of  concrete \coprob{} games}
In the game presented below, players walk on paths since it appears, in light of the literature, that such actions are most general. We also grant the cops a watch zone that enables them to capture the robbers whenever they are observed. We write $\pt{}$ for the set of finite paths in a graph and $\pt{v}\subseteq \pt{}$ for the set of paths that start on vertex $v\in V$. To simplify the notation, we formulate the concrete model in the setting where there are one cop and one robber, and without the auxiliary information set $\sto$. The extension to the general case is straightforward.

\begin{definition}\label{def:genconcretmodel}
A  GPCR game $\game=\left(S, \init, F, A, \tc, \tr \right)$  with  one cop and one robber (Definition~\ref{def:genabspurgame}) 
 is \emph{concrete} if there is a graph  $G=(V,E)$ satisfying:
\begin{enumerate}
\item $S = \stc \times \str $ is a finite set of configurations of the game.
\item $\init = (i_\cops,i_\robbers)$, where $i_\cops, i_\robbers\not\in V$.
\item $\stc \subseteq V\times \cl{P}(E) \cup\{ i_\cops\}$ is the set of configurations of the cop. 
%$V^k$ contains the positions of the $k$ cops and 
The second coordinate is the cop's  watch zone.
\item $\str \subseteq V\cup\{ i_\robbers\}$ is the set of positions of the robber. 
%\item $A = \ac \cup \ar$ is the finite set of actions of the cops and the robbers.
\item $\acs{(c,z),r} \subseteq \pt{c}\times \cl{P}(E)$ is the set of available actions for the cop. He  can move along a path from his present position $c$ and choose a watch~zone. From the initial state, $\acs{i_\cops} \subseteq V\times \cl{P}(E)$.
\item $\ars{(c,z),r} \subseteq \pt{r}$ is the set of available actions for the robber. She can move along a path from her present position $r$. From the initial state, $\ars{i_\robbers} \subseteq V$.
%\item $\tc : S\times \ac \times S \rightarrow [0,1]$ is a function satisfying $\sum_{s'\in S}\tc(s,a,s')\in \se{0,1}$ for all $s\in S$ and $a\in \ac$. Action $a$ can be played from state $s$ if and only if this value is $1$. This function is specific to the cops.
%\item $\tr(s,a,s')$ is a function similar to $\tc(s,a,s')$, but specific to the robbers.
\end{enumerate}
\end{definition}

{The definition of a play}  and all previous remarks and details that apply to Definition~\ref{def:genabspurgame} are still applicable in Definition~\ref{def:genconcretmodel}. 
%The cops form a team of $k$ agents against the $l$ robbers roaming on the graph. 
A peculiarity here is how we let the cop have his own watch zone consisting in a set of edges. Thus, the cop can only capture the robber on the robber's turn. Indeed, seeing as the robber moves along paths, we can explicitly deduce at what point a robber  is susceptible to get caught crossing a cop's watch zone. It's a natural choice of modeling that makes writing the probability of capture easier. 

\subsection{Example: Classic Cop and Robber game}\label{ex:classiccr}
Nowakowski and Winkler's, and Quilliot's, game is now presented  in the form of Definition \ref{def:genconcretmodel}. In this game, we will consider that the game is over not when the cop reaches the same position as the robber, but exactly after that, during the robber's turn, when she tries to escape. This slightly different interpretation leads to the same game. Our presentation allows to model and solve a more general situation where the robber could have a possibility of escaping, even if the cop reaches the robber's position. Let $G=(V,E)$ be a finite, undirected, reflexive and connected graph and let:
\begin{align*}
\stc &=  V\times \cl{P}(E)\\
\str &=  V\\
% The set of states is $S:=V^2$, we thus write $s=(c,r)$ for the positions of the cop and the robber. Actions of both players simply consist in  moving on adjacent vertices of their positions, so 
\acs{c,r}&= \{([c,c'],E_{c'})\mid [c,c']\in E \}.
\end{align*}
The watch zone $E_{c'}$ of the next state  is the set of adjacent edges of the cop's next position $c'$. The final states are those in which both players stand on the same vertex, $F=\se{(c,r)\in S : c=r}$. The initial state is  $\init=(\ic, \ir)$ and we let players choose any vertex from it, that is, $\acs{\ic, \ir} = \ars{c, \ir} = V$, with $c\in V$. Finally, the probabilities of transition are trivial since the game is deterministic. 

Now, in order to show that Equation \eqref{eq:absgamewn} is well-defined, we demonstrate how it encodes the relation $\preceq_n$ of Nowakowski and Winkler \cite{Nowakowski1983}. Since the game is deterministic, Equation \eqref{eq:absgamewn} reduces to :
\begin{align}\label{eq:wnclassic}
w_0(c,r) &=1 \iff c=r \nonumber\\
w_n(c,r) &= 
\max_{c'\in N[c]}\min_{r'\in N[r]}
w_{n-1}(c', r').
\end{align}
This equation is also a particular case of Equation~\eqref{eq:detergenabswn}. The next proposition shows that Equation \eqref{eq:wnclassic} simulates the relation $\preceq_n$. 
\begin{proposition}\label{prop:classwneq}
It holds that $w_n(c,r) = 1$ if and only if there exists a vertex $c'\in N[c]$ such that $r\preceq_n c'$.
\end{proposition}
\begin{proof}
We prove the result by induction. We note that in recursion $w_n$ it is the cop's turn to play, while in relation $\preceq_n$ the robber is next to move.
\\\textit{Base case: $n=0$.} $w_0(c,r)=1$ if and only if $r=c$ and $r=c$ if and only if $r\preceq_0 c$.
\\\textit{Induction step.} Assume the result holds for $n\leq k$ and let us show it holds for $n=k+1$. Then, $w_{k+1}(c,r)=1$ if and only if there exists an action $c'$ for the cop from which, no matter the response $r'$ of the robber, we have $w_k(c',r')=1$. By the induction hypothesis, $w_k(c',r')=1$ if and only if there exists a vertex $c''\in N[c']$ such that $r'\preceq_k c''$. Thus, the cop can play action $c'\in N[c]$ and we have $r\preceq_{k+1} c'$. Conversely, if there exists a vertex $c'\in N[c]$ such that $r\preceq_{k+1}c'$, then, by definition, for any action $r'\in N[r]$ of the robber there exist a response $c''\in N[c]$ of the cop such that $r'\preceq_k c''$. By the induction hypothesis, we thus have $w_k(c', r')=1$. In this case, the cop can play action $c'\in N[c]$ such that, no matter the answer of the robber $r'\in N[r]$, $w_k(c',r')=1$. By definition, we thus have $w_{k+1}(c,r)=1$.
\end{proof}

\subsection{Example: Cop and Fast Defending Robber game}
\label{ex:fastdefrobber}

Definition~\ref{def:genconcretmodel} is further illustrated on the following example. It describes the game of Cop and Fast Robber with probability of capture, which is a variant of the one presented by Fomin et al.~\cite{FominGKNS10}, already mentioned in Example~\ref{ex:copfastrobber}, and a variant of Example~\ref{ex:copdrunkdefrobber}, where the robber could evade from capture.
%Marcoux \cite{Marcoux} and 
%Chalopin et al.~\cite{Chalopin2011}. 
Unsurprisingly, given both games ask of robbers to move along paths, it is easier to write this new game following Definition~\ref{def:genconcretmodel}. 

For a path $\pi\in \pt{}$ on a graph $G$, we write $\pi[k]$ for its $k^{th}$ vertex and $\pi[*]$ for its last one.
Let $G=(V,E)$ be a finite graph. Assume that the cop guards a watch zone $C\subset E$ and that each time the robber crosses an edge $e$ he survives his walk with probability $q_C(e)$ (between 0 and 1). In Example~\ref{ex:copdrunkdefrobber}, a capture probability was used, here we define a survival probability as it is simpler to use in the current context. 
%We assume that $q_C(e)\geq 0$ for all edge $e$ and $q_C(e)<1$ whenever $e\in C$. 
Contrary to the Defending Robber game \ref{ex:copdrunkdefrobber}, the probability of survival depends on the cop's watch zone as well as the robber's action. Here, only the cop's watch zone and the transition functions are modified compared to Example~\ref{ex:copfastrobber}. So we have an element $\jail\notin V$ and the set of final states are $F=\se{(\jail, \emptyset, \jail)}$. 
We write $E_c$ for the set of edges incident to $c$. 
%We write $E_c^{\leq 1}$ for the set of edges inside a radius one of  a vertex $c$; that is, it is the set of edges linking .
 Similarly, we write $E_\pi$ for the set of edges of a path $\pi$. Let:
 %Concerning the game of cop and fast robber \ref{ex:copfastrobber}, the transition function of the cop simply sees its component $N[c]$ be replaced by $E_c$, that is, the cop now watches the edges contained in his neighborhood. The cop's transition function is:
\begin{align*}
\tc((c,E_c,r), c')
&=
\begin{cases}
\dirac{(c', E_{c'}, r)}, &\mbox{ if } c=\ic\mbox{ and } c'\in V \\
			&\mbox{ or if } c\in V \mbox{ and } c'\in N[c];\\
0, &\mbox{ otherwise.}
\end{cases}
\end{align*}
The robber's transition function is given by:
\begin{align*}
\tr((c,E_c, r), \pi) 
&=
\begin{cases}
\dirac{(c,E_c,\pi[*])}, &\mbox{ if } E_\pi \cap E_c =\emptyset;\\
D_{(r, \pi[*])}, &\mbox{ if } E_\pi \cap E_c \neq \emptyset;
\end{cases}
\end{align*}
where $D_{(r, \pi[*])}$ is a function satisfying:
\[
D_{(r, \pi[*])}(x) 
= 
\begin{cases}
\prod_{e\in E_\pi}q_{E_c}(e), &\mbox{ if } x=(c, E_c, \pi[*]);\\
1-\prod_{e\in E_\pi}q_{E_c}(e), &\mbox{ if } x=(\jail, \emptyset, \jail).
\end{cases}
\]
Note that to retrieve the game considered in Example~\ref{ex:copfastrobber} and Marcoux's thesis~\cite{Marcoux}, we should rather use a watch zone $E_c$ containing all edges on paths of length 2 from $c$ and change the conditions on $\tr$ for $E_{\pi_1}\cap E_c =\emptyset$ and $E_{\pi_1} \cap E_c \neq\emptyset$, where ${\pi_1}$ is the subpath of $\pi$ starting in $\pi[1]$.

Since the watch zone is determined by the cop's position, we can use the simplified notation $(c,r)$ for a state $(c,E_c,r)$. Thus, the recursion of Equation \eqref{eq:absgamewn} can be written as follows.
%\begin{align*}
%w_0&(c,r) = 1 \iff (c,E_c,r) = (\jail, \emptyset, \jail);\\
%w_n&(c,r) = \\
%&\begin{cases}
%1, &\mbox{ if } (c,E_c,r) = (\jail, \emptyset, \jail);\\
%\displaystyle
%\max_{c'\in N[c]}
%\min_{\pi\in\pt{r}}
%\tr\left((c', r), \pi, (c', \pi[*])\right)w_{n-1}(c', \pi[*])\hspace{-8mm}\mbox{~}&\\
%+\tr((c', r), \pi, (\jail, \emptyset, \jail)), &\mbox{ otherwise.}
%\end{cases}
%\end{align*}
For the jail state: $w_i (\jail, \emptyset, \jail) = 1$ for all $i\geq 0$. For $(c,E_c,r) \neq (\jail, \emptyset, \jail)$, we have $w_0(c,r) = 0$ and, for $n\geq 1$,
\begin{align*}
w_n(c,r)& = \\
\max_{\!\!\!\!c'\in N[c]}&
\min_{\pi\in\pt{r}\!\!}
\tr((c', r), \pi, (c', \pi[*]))w_{n-1}(c', \pi[*])+\tr((c', r), \pi, (\jail, \jail)).
\end{align*}

Following Proposition \ref{prop:compabswn}, the algorithmic complexity of the previous recursion is at most $\bigO{n\Delta\abs{V}^6\abs{\pt{}}}$, where $\Delta$ is the maximal degree of $G$. Indeed, $S$ corresponds to the set of pairs of vertices, the cop can only move on his neighbourhood and the robber is allowed to choose any path of finite length. Hence, even if we restrict the possible paths that can choose the robber to elementary paths (paths that do not cross twice a same vertex), the size of the possible robber actions, and therefore the size of $\cal{P}$, is exponential in the size of the graph on which the game is played. However, as shown in the next proposition, $w_n$ can be computed in polynomial time in the size of the graph itself.
\begin{proposition}
Computing $w_n(i)$ in the Cop and Fast Defending Robber game requires at most 
$\bigO{\abs{V}^3\log\abs{V} + (n+1)\abs{V}^2\abs{E}}$
%$\bigO{n\Delta\abs{V}^2}$ 
operations and uses at most $O({\abs{V}^3})$ space, for any $n\in \N$.
\end{proposition}
\begin{proof}
Let $\pth{r}{r'}$ be the set of paths beginning in $r$ and ending in $r'$. 
Let $(c,E_c)$ be a cop position. 
%The action $\pi$ of the robber can be decomposed on each of the edges $e$ of $E_\pi$. 
The robber's transition function can be simplified by assuming $q_{E_c}(e) = 1$ if $e\notin E_c$. Then, $\tr((c, r), \pi, (c,\pi[r']))= \Pi_{e\in E_\pi}q_{E_c}(e)$ if the robber is not caught on $\pi[r']$. The previous recursion, when state $(c,r)$ is not final, can be simplified to:
\begin{align*}
w_n(c,r) &=
\max_{c'\in N[c]}
\min_{\substack{r'\in V\\\pi\in \pth{r}{r'}}}
\left(
\prod_{e\in E_\pi} q_{E_{c'}}(e)	w_{n-1}(c',r')
+1 - \prod_{e\in E_\pi}
q_{E_{c'}}(e)
\right).
\\
&=\max_{c'\in N[c]}
\min_{\substack{r'\in V\\\pi\in \pth{r}{r'}}}
\Big(
(w_{n-1}(c',r') -1)\prod_{e\in E_\pi}  q_{E_{c'}}(e)	%w_{n-1}(c', v)
+1 \Big).
\end{align*}
%Exploring the set of neighbors of the cop position $c$ requires $\Delta$ operations. 
If $c'$ and $r'$ are fixed, we look for the path $\pi$ minimizing the expression in parentheses, so maximizing  $\prod_{e\in E_\pi}q_{E_{c'}}(e)$. This is  the same path that maximizes $\sum_{e\in E_\pi}\log q_{E_{c'}}(e)$ because $\log$ is a monotone increasing function. Because $\log q_{E_{c'}}(e)<0$, with $q_{E_{c'}}(e) \in [0,1]$, we can minimize $\sum_{e\in E_\pi} -\log q_{E_{c'}}(e)$. 
%We can thus consider the  shortest path from $r$ to any destination $r'$ by weighting each edge $e\in E$ with $-\log q_{E_c}(e)$. 

Observe that the survival probabilities $q_{E_{c'}}(e)$ depend only on the vertex $c'$ and the edge $e$. Thus, prior to evaluating $w_n(c,r)$, we can precompute $\abs{V}$ all-pairs shortest paths (one for each possible cop position) by weighting each edge $e\in E$ with $-\log q_{E_{x}}(e)$ for each source $x\in V$. This is done in $\bigO{\abs{E}\abs{V}^2 + \abs{V}^3\log\abs{V}}$ operations, for example by using the algorithm of Fredman and Tarjan \cite{Fredman1987}. This takes $\bigO{\abs{V}^3}$ space (because the path does not have to be stored, it can be recomputed in $\bigO{\Delta\abs{V}}$). Thus, for each $c'$ we store the values $\prod_{e\in E_\pi}  q_{E_{c'}}(e)$ for the shortest path $\pi$ between $r$ and $r'$. Finding the next robber position $r'$ thus requires at most $\bigO{\abs{V}}$ operations.

Now, assume $w_{n-1}(c',r')$ is computed for all $c'$, $r'$, in time $a_{n-1}$. We look for the vertex $c' \in N[c]$ maximizing
\[
\min_{\substack{r'\in V\\\pi\in \pth{r}{r'}}}
\Big((w_{n-1}(c',r') -1)\prod_{e\in E_\pi}  q_{E_{c'}}(e)	%w_{n-1}(c', v)
+1 \Big).
\] 
The values $w_{n-1}(c',r') $  are already computed, as well as the $\prod_{e\in E_\pi}  q_{E_{c'}}(e)$'s. Thus, at most $\bigO{\abs{N[c]}\abs{V}}$ operations are required to evaluate expression $w_n(c,r)$ when $c$ and $r$ are fixed. To find all maxima, that is for all $c\in V$ and $r\in V$, we need to make at most $\bigO{\abs{V}\sum_{c\in \abs{V}} \abs{N[c]}}=\bigO{2\abs{V}\abs{E}}$ operations. On turn $n$, we make a number of operations $a_n \in 
\bigO{\abs{V}\abs{E}} + a_{n-1}
\subseteq
\bigO{n\abs{V}\abs{E}}.
$
The total complexity is thus: 
\[
\bigO{\abs{E}\abs{V}^2 + \abs{V}^3\log\abs{V}} + a_n = \bigO{(n+1)\abs{E}\abs{V}^2 + \abs{V}^3\log\abs{V}}.
\] 
The bottlenecks for the spatial complexity are the shortest path algorithms which require at most $\bigO{\abs{V}^3}$ space. For each $w_n$ we only need $w_{n-1}$ so we do not need to store any other $w_k$ for $k<n-1$.
\end{proof}

An important aspect of the fast robber game is its ability to model situations of imperfect information in which the cops only gather information on the robber's position at regular intervals. This game, deemed with \emph{witness}, is shown by Chalopin et al.~\cite{Chalopin2011} to correspond to the game of Cop and Fast Robber (without watch zone). In essence, the authors present an equivalence between the classes of \win{} graphs in the witness game and in the fast robber one. We can wonder if the same could be said of the stochastic case.

\subsection{Example: Cop and Drunk Robber game}
Let us revisit the Cop and Drunk Robber game of Example~\ref{ex:copdrunkrobber} with the concrete model and Equation~\eqref{eq:absgamewn}.

We can show how it is always easier to capture a robber moving randomly than a robber playing optimally. For the sake of generality, assume the robber can play according to any distribution in $\dist{N[r]}$ when she finds herself on vertex $r$. Let $\phi\subseteq\left(\dist{N[r]}\right)_{r\in V}$ be a sequence of distributions on the vertices $V$ and $\phi_r$ its component that is in $\dist{N[r]}$. Then, we write $w_n^\phi(c,r)$ for the recursion in which $\tr((c,r), r') = \phi_r$. In other words, we write:
\begin{align}\label{eq:randomrobberwn}
w_n^\phi(c,r) = \max_{c'\in N[c]}\sum_{r'\in N[r]}\phi_r(r')w_{n-1}^\phi(c', r'),
\end{align}
if $c\neq r$ and $w_n^\phi(c,r)=1$ if $c=r$, for all $n\geq 0$. The classic recursion from Equation \eqref{eq:wnclassic} is written $w_n(c,r)$.
\begin{proposition}
It is always easier to capture a robber playing randomly than an adversarial one, that is:
\[
w_n^\phi(c,r) \geq w_n(c,r).
\]
\end{proposition}
\begin{proof}
We write $\dirac{N[r]}$ for the set of Dirac distributions defined on $N[r]$. The robber would be harder to capture if she were to minimize her probability of capture only on Dirac distributions because her optimal strategy is deterministic. Thus, we compute
\begin{align*}
w_n^\phi(c,r) &:= 
\max_{c'\in N[c]}\sum_{r'\in N[r]}\phi_r(r')w_{n-1}^\phi(c', r')\\
&\geq 
\max_{c'\in N[c]}\min_{\psi\subseteq \left(\dist{N[r]}\right)_{r\in V}}\sum_{r'\in N[r]}\psi_r(r')w_{n-1}^\psi(c', r')\\
&=\max_{c'\in N[c]}\min_{\psi\subseteq \left(\dirac{N[r]}\right)_{r\in V}}\sum_{r'\in N[r]}\psi_r(r')w_{n-1}^\psi(c', r')\\
&=\max_{c'\in N[c]}\min_{r'\in N[r]}w_{n-1}(c', r').
\end{align*}
The first line is the definition of Equation \eqref{eq:randomrobberwn} with the robber playing according to distribution $\phi$. If she could choose this distribution, she could fall on a distribution $\psi\subseteq \left(\dist{N[r]}\right)_{r\in V}$ ensuring her a greater probability of survival, that justifies the second line. Then, we observe that since her optimal strategy is deterministic, it corresponds to a sequence of Dirac distributions and she loses nothing in playing according to $\psi\subseteq \left(\dirac{N[r]}\right)_{r\in V}$. The last line is simply the preceding one rewritten without distributions, as in this case $\psi_r$ is concentrated on a single vertex $r'\in N[r]$.
\end{proof}

% \subsection{Example: The game on a dynamic graph}\label{ex:dynamicgraphgame}
\subsection{Example: Temporal Cop and Robber game}\label{ex:dynamicgraphgame}
In graph theory, one can define many random processes to stochastically generate graphs that vary on each time step. One thus obtains a sequence of graphs $G_0, G_1, \dots$ that represents the evolution of a network over time. Those graphs are called \emph{dynamic graphs, link streams, time-varying graphs or temporal networks}, depending on the community, and can model, for example, the destruction of a bridge or of a road that makes it impossible for the players to pass through it.

Suppose $k$ cops are chasing $l$ robbers on the sequence $G_0, G_1, \dots$. In order to take into account the variable nature of the underlying structure of a game from Definition \ref{def:genconcretmodel}, we can make use of the component $\sto$ as a turn counter. Let $G_t=(V_t, E_t)$ be the graph generated at time $t$,
\[
S_t = V_t^k \times \cl{P}(E_t)^k \times V_t^l \times \se{t}
\]
and $S = \bigcup_{t=1}^\infty S_t$. Hence, on each time step $t$ a new graph $G_t$ is created according to a certain process and the set of states is renewed. The sets of actions can also be redefined. Let $\pth{u}{G_t}$ be the set of finite paths on $G_t$ that begin on vertex $u\in V_t$. The sets of actions are thus :
\begin{align*}
\acs{c,C,r,t} &\subseteq \prod_{i=1}^k \pth{c^i}{G_t}\times \cl{P}(E_t);\\
\ars{c,C,r,t} &\subseteq \prod_{i=1}^l \pth{r^i}{G_t}.
\end{align*}
Since this example is rather general, we let the transition functions be undefined. We require, however, that the transition functions follow the arrow of time: if $s_t\in S_t$ is a game state at time $t$ and $\cac\in \acs{s_t}$ is a cop action, then $\tc(s_t,\cac)(X)>0$ only if $X\subseteq S_{t+1}$. The same holds for the robbers.

%!TEX root = ms.tex

\section{Conclusion}\label{sec:conclusion}
This paper presented a relatively simple yet very general model in order to describe games of \coprob{} that, notably, may include stochastic aspects. 
 The game $\game$ was presented along with a method of resolution in the form of a recursion $w_n$ in Theorem~\ref{thm:copwinthm}. 
We show in Proposition~\ref{prop:epsoptimal} that we can always retrieve an $\epsilon$-optimal strategy for $\game$ from the recursion $w_n$ (for large enough $n$). 
Moreover, in Proposition \ref{prop:wnstation} we show that if the recursion becomes stationary, stationarity must occur at most at index $\abs{S}$. This is a first step in the analysis of the rate of convergence of the recursion.

We have exposed how some classic Cops and Robbers games can be written into our model and 
%in some cases, 
extended. Many more games could now be studied as GPCR such as the Firefighting game, under certain conditions, in which a team of firefighters seeks to prevent the nodes of a graph from burning.
 An interesting notion that is captured by our framework, in Definition~\ref{def:genconcretmodel}, is that of the surveillance zones of the cops that can be chosen at each step. 
Thus, we claim a wide variety of games of \coprob{} can be solved with the concepts developed in this paper. Furthermore, such a broad exposition of games of \coprob{} as ours enables one to study the effects of modifying certain rules, for example on the number of cops or on the speed of players, on the games. That is, one can use Equation~\eqref{eq:absgamewn} and probe its values in order to test how modifying these rules affect the ability of the cops to capture the robbers. %In other words, one could study under what types of rule changes would the \emph{\win{}} property of a game remain. 
%Many of our results have an algorithmic flavour, which allows concrete implementation. For example, 

We have extended the classic notion of cop number with the \pcoptext{}, although the question is still open about the behaviour of this function. 
The expected capture time of the robbers is also of great interest. This function can now be studied on large swaths of \coprob{} games. In part, this question can be motivated by a Simard et al.'s paper~\cite{Simard2015} on the relation between an Operations Research problem and the resolution of a Cop and Drunk Robber game. Specifically, the authors tackled the problem of upper bounding the probability of detecting a hidden and randomly moving object on a graph with a single optimally moving searcher. This problem, being \NPH{} ~\cite{Trummel}, is constrained to be solved in a maximum number of time steps $T\in \N$. In particular, it appears that if one could tightly upper bound the expected capture time of a game derived from Definition~\ref{def:genabspurgame}, then one could, following the ideas presented in this paper, deduce the optimal number of searchers to send on a mission to rescue the object. Then, if this number were deduced, one could further apply the ideas of this article along with Equation~\eqref{eq:absgamewn} in order to help solve this search problem with multiple searchers. %This is only a small justification for studying those questions, albeit, it seems, a valuable one.

Finally, a last avenue of research that is worth mentioning and that is possibly of most interest to researchers in robotics and operations research concerns the extension of model~\ref{def:genabspurgame} to games of imperfect information. Imperfect information refers to the lack of knowledge of one or both players. Cops and Robbers games of imperfect information thus contain games in which robbers are invisible, that can model problems of graph search such as the one mentioned above. Game theory seems apt to enable the transition from perfect information to imperfect information games with the use of belief states. Such generalization could be paired with the \emph{branch and bound} method presented in Simard et al.~\cite{Simard2015} in order to solve more general search problems.

In light of the literature on \coprob{} games it appears this paper distances itself from most studies on the subject. Indeed, we do not claim any results on typical \coprob{} questions such as the asymptotic behaviour of $\pcopsym$ or on dismantling schemes to characterize classes of winning graphs. However, we think that modelling such a wide variety of games opens the door to further studies on \coprob{} games that can now be tackled in their generality, which was not possible before. Thus, although our model may not enable one to compute analytical solutions on classical questions of \coprob{} games, we have good hope that algorithmic ones will be devised in order to solve more general problems on classes, not of graphs, but of games. In short, it appears that new and promising avenues of research have come to light with the objects presented in this paper and we hope researchers will be driven to tackle those open questions that were unearthed.

\section*{Acknowledgement}
 The authors acknowledge the careful reading of reviewers, which has helped improve the paper presentation. Josée Desharnais and François Laviolette  acknowledge the support of the Natural Sciences and Engineering Research Council of Canada (NSERC, grant numbers 239294 and 262067).

% \section*{References}
\bibliographystyle{plain}
\bibliography{Copwin}

\begin{thebibliography}{10}

\bibitem{Aigner1984a}
M.~Aigner and M.~Fromme.
\newblock {A game of cops and robbers}.
\newblock {\em Discrete Applied Mathematics}, 8(1):1--12, 1984.

\bibitem{Bonatoa}
Anthony Bonato and Gary Macgillivray.
\newblock {Characterizations and algorithms for generalized cops and robbers
  games}.
\newblock {\em Contributions to Discrete Mathematics}, 12(1):1--10, 2017.

\bibitem{Bonato2016APV}
Anthony Bonato, Dieter Mitsche, Xavier P{\'e}rez-Gim{\'e}nez, and Pawel Pralat.
\newblock A probabilistic version of the game of zombies and survivors on
  graphs.
\newblock {\em Theor. Comput. Sci.}, 655:2--14, 2016.

\bibitem{Bonato2011f}
Anthony Bonato and Richard~J. Nowakowski.
\newblock {\em {The Game of Cops and Robbers on Graphs}}.
\newblock American Mathematical Society, 2011.

\bibitem{boyer2013cops}
M~Boyer, S~El~Harti, A~El~Ouarari, R~Ganian, T~Gavenciak, G~Hahn, C~Moldenauer,
  I~Rutter, B~Th{\'e}riault, and M~Vatshelle.
\newblock Cops-and-robbers: remarks and problems.
\newblock {\em Journal of Combinatorial Mathematics and Combinatorial
  Computing}, 85, 2013.

\bibitem{Chalopin2011}
J{\'{e}}r{\'{e}}mie Chalopin, Victor Chepoi, Nicolas Nisse, and Yann
  Vax{\`{e}}s.
\newblock {Cop and Robber Games When the Robber Can Hide and Ride}.
\newblock {\em SIAM Journal on Discrete Mathematics}, 25(1):333--359, jan 2011.

\bibitem{Clarke2012}
Nancy~E. Clarke and Gary MacGillivray.
\newblock {Characterizations of k-copwin graphs}.
\newblock {\em Discrete Mathematics}, 312(8):1421--1425, 2012.

\bibitem{condon1992complexity}
Anne Condon.
\newblock The complexity of stochastic games.
\newblock {\em Information and Computation}, 96(2):203--224, 1992.

\bibitem{Conway1976}
John~Horton Conway.
\newblock {\em {On numbers and games}}.
\newblock IMA, 1976.

\bibitem{FominGKNS10}
Fedor~V. Fomin, Petr~A. Golovach, Jan Kratochv{\'{\i}}l, Nicolas Nisse, and
  Karol Suchan.
\newblock Pursuing a fast robber on a graph.
\newblock {\em Theoretical Computer Science}, 411(7-9):1167--1181, feb 2010.

\bibitem{Fomin2008}
Fedor~V. Fomin and Dimitrios~M. Thilikos.
\newblock {An annotated bibliography on guaranteed graph searching}.
\newblock {\em Theoretical Computer Science}, 399(3):236--245, 2008.

\bibitem{Fredman1987}
Michael~L. Fredman and Robert~Endre Tarjan.
\newblock {Fibonacci heaps and their uses in improved network optimization
  algorithms}.
\newblock {\em Journal of the ACM}, 34(3):596--615, 1987.

\bibitem{gimbert2008simple}
Hugo Gimbert and Florian Horn.
\newblock Simple stochastic games with few random vertices are easy to solve.
\newblock In {\em International Conference on Foundations of Software Science
  and Computational Structures}, pages 5--19. Springer, 2008.

\bibitem{granas2003}
Andrzej Granas and James Dugundji.
\newblock {\em Fixed Point Theory}.
\newblock Springer-Verlag, 2003.

\bibitem{Hahn2006}
Ge{\v n}a Hahn and Gary MacGillivray.
\newblock A note on k-cop, l-robber games on graphs.
\newblock {\em Discrete Mathematics}, 306(19):2492 -- 2497, 2006.
\newblock Creation and Recreation: A Tribute to the Memory of Claude Berge.

\bibitem{Kehagias2018}
Ath. Kehagias.
\newblock Generalized cops and robbers: A multi-player pursuit game on graphs.
\newblock {\em Dynamic Games and Applications}, Nov 2018.

\bibitem{KEHAGIAS201725}
Ath. Kehagias and G.~Konstantinidis.
\newblock Selfish cops and passive robber: Qualitative games.
\newblock {\em Theoretical Computer Science}, 680:25 -- 35, 2017.

\bibitem{KEHAGIAS2013100}
Athanasios Kehagias, Dieter Mitsche, and Paweł Prałat.
\newblock Cops and invisible robbers: The cost of drunkenness.
\newblock {\em Theoretical Computer Science}, 481:100 -- 120, 2013.

\bibitem{Kehagias2012}
Athanasios Kehagias and Pawe{\l} Pra{\l}at.
\newblock Some remarks on cops and drunk robbers.
\newblock {\em Theoretical Computer Science}, 463:133 -- 147, 2012.
\newblock Special Issue on Theory and Applications of Graph Searching Problems.

\bibitem{Kinnersley2015}
William~B. Kinnersley.
\newblock {Cops and Robbers is EXPTIME-complete}.
\newblock {\em Journal of Combinatorial Theory, Series B}, 111:201--220, mar
  2015.

\bibitem{Komarov}
Natasha Komarov.
\newblock {\em {Expected Capture Time in Variants of Cops {\&} Robbers Games}}.
\newblock PhD thesis, Dartmouth College, 2013.

\bibitem{Komarov2013b}
Natasha Komarov and Peter Winkler.
\newblock {Capturing the Drunk Robber on a Graph}.
\newblock {\em The Electronic Journal of Combinatorics}, 21(3):14, 2014.

\bibitem{Konstantinidis2017}
G.~Konstantinidis and A.~Kehagias.
\newblock Selfish cops and active robber: Multi-player pursuit evasion on
  graphs.
\newblock {\em Theoretical Computer Science}, 780:84 -- 102, 2019.

\bibitem{KONSTANTINIDIS201648}
G.~Konstantinidis and Ath. Kehagias.
\newblock Simultaneously moving cops and robbers.
\newblock {\em Theoretical Computer Science}, 645:48 -- 59, 2016.

\bibitem{Marcoux}
H{\'{e}}li Marcoux.
\newblock {\em {Jeux de poursuite policier-voleur sur un graphe, le cas du
  voleur rapide}}.
\newblock M{\'{e}}moire de ma{\^{i}}trise, Universit{\'{e}} Laval, 2014.

\bibitem{Nowakowski1983}
Richard Nowakowski and Peter Winkler.
\newblock {Vertex-to-vertex pursuit in a graph}.
\newblock {\em Discrete Mathematics}, 43(2-3):235--239, 1983.

\bibitem{osborne1994course}
Martin~J Osborne and Ariel Rubinstein.
\newblock {\em A course in game theory}.
\newblock MIT press, 1994.

\bibitem{Puterman2014}
Martin~L. Puterman.
\newblock {\em {Markov decision processes: discrete stochastic dynamic
  programming}}.
\newblock John Wiley {\&} Sons, 2014.

\bibitem{Quilliot1978f}
Alain Quilliot.
\newblock {\em {Probl{\`{e}}mes de jeux, de point fixe, de connectivit{\'{e}}
  et de repr{\'{e}}sentation sur des graphes, des ensembles ordonn{\'{e}}s et
  des hypergraphes}}.
\newblock Th{\`{e}}se de doctorat d'{\'{e}}tat, Universit{\'{e}} de Paris VI,
  France, 1983.

\bibitem{shapley1953stochastic}
Lloyd~S Shapley.
\newblock Stochastic games.
\newblock {\em Proceedings of the national academy of sciences},
  39(10):1095--1100, 1953.

\bibitem{Simard2015}
Fr{\'{e}}d{\'{e}}ric Simard, Michael Morin, Claude-Guy Quimper, Fran{\c{c}}ois
  Laviolette, and Jos{\'{e}}e Desharnais.
\newblock {Bounding an Optimal Search Path with a Game of Cop and Robber on
  Graphs}.
\newblock In {\em Principles and Practice of Constraint Programming: 21st
  International Conference, CP 2015, Cork, Ireland, August 31--September 4,
  2015, Proceedings}, volume 9255, pages 403--418. Springer Science Business
  Media, 2015.

\bibitem{Trummel}
K~E Trummel and J~R Weisinger.
\newblock {The Complexity of the Optimal Searcher Path Problem.}
\newblock {\em Operations Research}, 34(2):324--327, 1986.

\bibitem{markovGames}
J.~{Wal, van der} and J.~Wessels.
\newblock {\em On Markov games}.
\newblock Memorandum COSOR. Technische Hogeschool Eindhoven, 1975.

\end{thebibliography}

\begin{appendices}
%!TEX root = ms.tex

\section{Constructing a GPCR game as a Simple Stochastic Game}\label{sec:annex_ssg}
The following argument is inspired by the SSG exposition of Gimbert and Horn \cite{gimbert2008simple}. A simple stochastic game is a tuple $(V, V_{\max}, V_{\min}, V_R, E, t, p)$, where $(V,E)$ describes a directed graph $G$ and $V_{\max}, V_{\min}, V_R$ form a partition of $V$. There is a special vertex $t\in V$, called the target, and $p$ is a probability function such that for every vertex $w\in V$ and $v\in V_R$, $p(w\mid v)$ is the probability of transiting from $v$ to $w$. There are two players, $\max$ and $\min$, and the game is played with perfect information. The set $V_{\max}$ contains those nodes controlled by player $\max$, i.e. where this player is next to play, and $V_{\min}$ those nodes controlled by player $\min$. The set of edges $E$ is defined by the possible moves in the game. The game proceeds as follows: imagine a token is placed on some initial vertex $i\in V_{\max}\cup V_{\min}$, then the player who is next to play moves the token along an edge, either the token is again in some vertex of $V_{\max}\cup V_{\min}$ where one player has to make a move, or the token is now on some vertex $v$ of $V_R$. When the token is on $v$, an outneighbour of $v$ is chosen randomly according to the distribution $p(\cdot\mid v)$, where the token is moved. The game ends if $t$ is ever encountered, in which case $\max$ wins, otherwise it continues indefinitely and the other player wins.

Following Gimbert and Horn, we define a play as an infinite sequence of vertices $v_0v_1\dots$ of $G$ such that $(v_i,v_{i+1})\in E$ for all $i$ and a finite play (what we called a history) as a finite prefix of a play. A strategy for $\max$ is a function $\sigma : V^*V_{\max}\rightarrow V$ and a strategy for $\min$ is a function $\tau : V^*V_{\min} \rightarrow V$, where $V^*$ is the set of finite plays. We suppose that for each finite play $(v_0\dots v_n)$ and vertex $v\in V_{\max}$, $(v,\sigma(v_0\dots v_n v)\in E$ and similarly for $\tau$. Note that such strategies are deterministic, which is without loss of generality. We write for convenience $\Gamma_{\max}$ and $\Gamma_{\min}$ for the sets of $\max$ (resp. $\min$) strategies. Now, for any node $v\in V$, we can define the value of $v$ for $\max$ (resp. $\min$) as the probability the target node is reached from that node. If $p(t \mid \sigma, \tau, v)$ is the probability that $t$ is reached from $v$ under strategies $\sigma$ and $\tau$, then we let 
\begin{align*}
\underline{val}(v) 
&:= \sup_{\sigma\in \Gamma_{\max}}\inf_{\tau\in\Gamma_{\min}}
p(t\mid \sigma,\tau,v),\\
\overline{val}(v)
&:=\inf_{\tau\in\Gamma_{\min}}\sup_{\sigma\in \Gamma_{\max}}
p(t\mid \sigma,\tau,v).
\end{align*} 

The following theorem \cite{condon1992complexity,gimbert2008simple,shapley1953stochastic} is well known about simple stochastic games.
\begin{theorem}\label{thm:ssgvalexist}
In any simple stochastic game and from any vertex $v$, $\underline{val}(v) = \overline{val}(v)$ and we write $val(v):= \underline{val}(v)$. Furthermore, there exists deterministic and memoryless strategies for players $\max$ and $\min$ that achieve the value $val(v)$.
\end{theorem}

We write a GPCR game $\game$ as a simple stochastic game by describing a directed graph $G = (V = V_\cops\cup V_\robbers\cup V_R\cup\se{t}, E)$, where $V_\cops$ is the set of vertices controlled by the cops, $V_\robbers$ the set of vertices controlled by the robbers, $V_R$ is the set of random vertices and $t$ is the target vertex for the cops. The set of edges $E$ is induced by the transition functions $\tc$ and $\tr$. If there exists a play in $\game$ with a subsequence $sas'$, then we add an edge from $s$ to some node $v\in V_R$, labelled $a$. We add an edge from $a$ to $s'$ weighted either by $\tc(s,a,s')$ or $\tr(s,a,s')$ depending on whether $a$ was played by the cops or by the robbers. We assume that vertex $t$ holds all final states of $F$, thus all transitions of the form $\tc(s,a,f)>0$ or $\tr(s,a,f)>0$ for any state $s$, action $a$ and final state $f$, induce edges from some vertex of $V_R$ to $t$. Now, in this game the cops win if and only if they can reach $t$ from the initial vertex $\init$. Thus, this is a simple stochastic game that corresponds to $\game$. We deduce from \autoref{thm:ssgvalexist} that $val(v)$ exists and it is the probability that the cops capture the robbers from a vertex, or state, $v$ in $\game$. Since this SSG has a value, so does $\game$ and this value is the probability just mentioned.

\end{appendices}

\end{document}